\documentclass[a4paper,12pt]{article}

\usepackage[top=1.2in,right=0.8in,left=.8in,bottom=1.2in]{geometry}

\usepackage{latexsym}
\usepackage{amssymb}
\usepackage{float}
\usepackage{graphics,graphicx}
\usepackage{tabularx}
\usepackage{amsfonts}
\usepackage{amsmath}
\usepackage{enumerate}
\usepackage{booktabs}
\usepackage{multirow}
\usepackage{sectsty}
\usepackage[affil-it,auth-sc]{authblk}
\usepackage{subfig}
\usepackage{color}
\usepackage{mathtools}
\usepackage{mathrsfs}
\usepackage{bmpsize}
\usepackage{xcolor}
\usepackage{siunitx}
\usepackage{hyperref}
\usepackage{xr}
\usepackage{longtable}

\usepackage{multirow}
\usepackage{array}
\usepackage{tabu}
\usepackage{mathtools}
\usepackage{hyperref}
\usepackage{cancel}

\usepackage[normalem]{ulem}               
\newcommand\redout{\bgroup\markoverwith
	{\textcolor{red}{\rule[0.5ex]{2pt}{0.8pt}}}\ULon}
\usepackage{amsmath}
\usepackage{amsfonts}
\usepackage{amssymb}
\usepackage{siunitx} 

\usepackage{enumerate}
\usepackage{sectsty}
\usepackage[affil-it,auth-sc]{authblk}
\usepackage{color}
\sectionfont{\normalsize}
\subsectionfont{\normalsize}

\usepackage{fancyhdr}            
\fancypagestyle{plain}{%
  \fancyhead{}                                     
  \fancyfoot[CE,CO]{\thepage}       
  \fancyhead[CE,CO]{\textcolor[rgb]{0.64,0.15,0.15}{Published in \emph{Journal of the Mechanics and Physics of Solids} (2020) \textbf{}
  \\
doi: \href{https://doi.org/10.1016/j.jmps.2019.103735}{10.1016/j.jmps.2019.103735}
}}
\setlength{\headheight}{14.5pt}}

\newcommand{\footmsg}[1]{%
  \let\temp\thempfn%
  \def\thempfs{}
  \footnotetext{#1}
  \let\tempfn\temp}

\begin{document}

\newcommand{\singlespace}{\baselineskip=12pt\lineskiplimit=0pt\lineskip=0pt}
\def\ds{\displaystyle}

\newcommand{\beq}{\begin{equation}}
\newcommand{\eeq}{\end{equation}}
\newcommand{\lb}{\label}
\newcommand{\ph}{\phantom}
\newcommand{\beqar}{\begin{eqnarray}}
\newcommand{\eeqar}{\end{eqnarray}}
\newcommand{\barr}{\begin{array}}
\newcommand{\earr}{\end{array}}
\newcommand{\jump}{\parallel}
\newcommand{\Ehat}{\hat{E}}
\newcommand{\That}{\hat{\bf T}}
\newcommand{\Ahat}{\hat{A}}
\newcommand{\chat}{\hat{c}}
\newcommand{\shat}{\hat{s}}
\newcommand{\khat}{\hat{k}}
\newcommand{\muhat}{\hat{\mu}}
\newcommand{\mc}{M^{\scriptscriptstyle C}}
\newcommand{\mei}{M^{\scriptscriptstyle M,EI}}
\newcommand{\mec}{M^{\scriptscriptstyle M,EC}}
\newcommand{\hbeta}{{\hat{\beta}}}
\newcommand{\rec}[2]{\left( #1 #2 \ds{\frac{1}{#1}}\right)}
\newcommand{\rep}[2]{\left( {#1}^2 #2 \ds{\frac{1}{{#1}^2}}\right)}
\newcommand{\derp}[2]{\ds{\frac {\partial #1}{\partial #2}}}
\newcommand{\derpn}[3]{\ds{\frac {\partial^{#3}#1}{\partial #2^{#3}}}}
\newcommand{\dert}[2]{\ds{\frac {d #1}{d #2}}}
\newcommand{\dertn}[3]{\ds{\frac {d^{#3} #1}{d #2^{#3}}}}
\newcommand{\ct}{\captionof{table}}
\newcommand{\cf}{\captionof{figure}}

\def\c{{\circ}}
\def\bob{{\, \underline{\overline{\otimes}} \,}}
\def\ob{{\, \underline{\otimes} \,}}
\def\scalp{\mbox{\boldmath$\, \cdot \, $}}
\def\gdp{\makebox{\raisebox{-.215ex}{$\Box$}\hspace{-.778em}$\times$}}
\def\daa{\makebox{\raisebox{-.050ex}{$-$}\hspace{-.550em}$: ~$}}
\def\mK{\mbox{${\mathcal{K}}$}}
\def\cK{\mbox{${\mathbb {K}}$}}

\def\Xint#1{\mathchoice
   {\XXint\displaystyle\textstyle{#1}}%
   {\XXint\textstyle\scriptstyle{#1}}%
   {\XXint\scriptstyle\scriptscriptstyle{#1}}%
   {\XXint\scriptscriptstyle\scriptscriptstyle{#1}}%
   \!\int}
\def\XXint#1#2#3{{\setbox0=\hbox{$#1{#2#3}{\int}$}
     \vcenter{\hbox{$#2#3$}}\kern-.5\wd0}}
\def\ddashint{\Xint=}
\def\fpint{\Xint=}
\def\dashint{\Xint-}
\def\cpvint{\Xint-}
\def\intl{\int\limits}
\def\cpvintl{\cpvint\limits}
\def\fpintl{\fpint\limits}
\def\ointl{\oint\limits}
\def\bA{{\bf A}}
\def\ba{{\bf a}}
\def\bB{{\bf B}}
\def\bb{{\bf b}}
\def\bc{{\bf c}}
\def\bC{{\bf C}}
\def\bD{{\bf D}}
\def\bE{{\bf E}}
\def\be{{\bf e}}
\def\bbf{{\bf f}}
\def\bF{{\bf F}}
\def\bG{{\bf G}}
\def\bg{{\bf g}}
\def\bi{{\bf i}}
\def\bH{{\bf H}}
\def\bK{{\bf K}}
\def\bL{{\bf L}}
\def\bM{{\bf M}}
\def\bN{{\bf N}}
\def\bn{{\bf n}}
\def\bm{{\bf m}}
\def\b0{{\bf 0}}
\def\bo{{\bf o}}
\def\bX{{\bf X}}
\def\bx{{\bf x}}
\def\bP{{\bf P}}
\def\bp{{\bf p}}
\def\bQ{{\bf Q}}
\def\bq{{\bf q}}
\def\bR{{\bf R}}
\def\bS{{\bf S}}
\def\bs{{\bf s}}
\def\bT{{\bf T}}
\def\bt{{\bf t}}
\def\bU{{\bf U}}
\def\bu{{\bf u}}
\def\bv{{\bf v}}
\def\bw{{\bf w}}
\def\bW{{\bf W}}
\def\by{{\bf y}}
\def\bz{{\bf z}}
\def\T{{\bf T}}
\def\Te{\textrm{T}}
\def\Id{{\bf I}}
\def\bxi{\mbox{\boldmath${\xi}$}}
\def\balpha{\mbox{\boldmath${\alpha}$}}
\def\bbeta{\mbox{\boldmath${\beta}$}}
\def\bepsilon{\mbox{\boldmath${\epsilon}$}}
\def\bvarepsilon{\mbox{\boldmath${\varepsilon}$}}
\def\bomega{\mbox{\boldmath${\omega}$}}
\def\bphi{\mbox{\boldmath${\phi}$}}
\def\bsigma{\mbox{\boldmath${\sigma}$}}
\def\bfeta{\mbox{\boldmath${\eta}$}}
\def\bDelta{\mbox{\boldmath${\Delta}$}}
\def\btau{\mbox{\boldmath $\tau$}}
\def\tr{{\rm tr}}
\def\dev{{\rm dev}}
\def\div{{\rm div}}
\def\Div{{\rm Div}}
\def\Grad{{\rm Grad}}
\def\grad{{\rm grad}}
\def\Lin{{\rm Lin}}
\def\Sym{{\rm Sym}}
\def\Skw{{\rm Skew}}
\def\abs{{\rm abs}}
\def\Re{{\rm Re}}
\def\Im{{\rm Im}}
\def\capB{\mbox{\boldmath${\mathsf B}$}}
\def\capC{\mbox{\boldmath${\mathsf C}$}}
\def\capD{\mbox{\boldmath${\mathsf D}$}}
\def\capE{\mbox{\boldmath${\mathsf E}$}}
\def\capG{\mbox{\boldmath${\mathsf G}$}}
\def\tcapG{\tilde{\capG}}
\def\capH{\mbox{\boldmath${\mathsf H}$}}
\def\capK{\mbox{\boldmath${\mathsf K}$}}
\def\capL{\mbox{\boldmath${\mathsf L}$}}
\def\capM{\mbox{\boldmath${\mathsf M}$}}
\def\capR{\mbox{\boldmath${\mathsf R}$}}
\def\capW{\mbox{\boldmath${\mathsf W}$}}

\def\i{\mbox{${\mathrm i}$}}
\def\mC{\mbox{\boldmath${\mathcal C}$}}
\def\mB{\mbox{${\mathcal B}$}}
\def\mE{\mbox{${\mathcal{E}}$}}
\def\mL{\mbox{${\mathcal{L}}$}}
\def\mK{\mbox{${\mathcal{K}}$}}
\def\mV{\mbox{${\mathcal{V}}$}}
\def\C{\mbox{\boldmath${\mathcal C}$}}
\def\E{\mbox{\boldmath${\mathcal E}$}}

\def\AAM{{\it Advances in Applied Mechanics }}
\def\ACME{{\it Arch. Comput. Meth. Engng.}}
\def\ARMA{{\it Arch. Rat. Mech. Analysis}}
\def\AMR{{\it Appl. Mech. Rev.}}
\def\ASCEEM{{\it ASCE J. Eng. Mech.}}
\def\ACTA{{\it Acta Mater.}}
\def\CMAME {{\it Comput. Meth. Appl. Mech. Engrg.}}
\def\CRAS{{\it C. R. Acad. Sci. Paris}}
\def\CRM{{\it Comptes Rendus M\'ecanique}}
\def\EFM{{\it Eng. Fracture Mechanics}}
\def\EJMA{{\it Eur.~J.~Mechanics-A/Solids}}
\def\IJES{{\it Int. J. Eng. Sci.}}
\def\IJF{{\it Int. J. Fracture}}
\def\IJMS{{\it Int. J. Mech. Sci.}}
\def\IJNAMG{{\it Int. J. Numer. Anal. Meth. Geomech.}}
\def\IJP{{\it Int. J. Plasticity}}
\def\IJSS{{\it Int. J. Solids Structures}}
\def\IngA{{\it Ing. Archiv}}
\def\JAM{{\it J. Appl. Mech.}}
\def\JAP{{\it J. Appl. Phys.}}
\def\JAE{{\it J. Aerospace Eng.}}
\def\JE{{\it J. Elasticity}}
\def\JM{{\it J. de M\'ecanique}}
\def\JMPS{{\it J. Mech. Phys. Solids}}
\def\JSV{{\it J. Sound and Vibration}}
\def\MACRO{{\it Macromolecules}}
\def\MMT{{\it Mech. Mach. Th.}}
\def\MOM{{\it Mech. Materials}}
\def\MMS{{\it Math. Mech. Solids}}
\def\MMT{{\it Metall. Mater. Trans. A}}
\def\MPCPS{{\it Math. Proc. Camb. Phil. Soc.}}
\def\MSE{{\it Mater. Sci. Eng.}}
\def\NATURE{{\it Nature}}
\def\NATUREM{{\it Nature Mater.}}
\def\PHIL{{\it Phil. Trans. R. Soc.}}
\def\PMPS{{\it Proc. Math. Phys. Soc.}}
\def\PNAS{{\it Proc. Nat. Acad. Sci.}}
\def\PRE{{\it Phys. Rev. E}}
\def\PRL{{\it Phys. Rev. Letters}}
\def\PRSL{{\it Proc. R. Soc.}}
\def\RIIT{{\it Rozprawy Inzynierskie - Engineering Transactions}}
\def\ROCK{{\it Rock Mech. and Rock Eng.}}
\def\QAM{{\it Quart. Appl. Math.}}
\def\QJMAM{{\it Quart. J. Mech. Appl. Math.}}
\def\SCIENCE{{\it Science}}
\def\SCRMAT{{\it Scripta Mater.}}
\def\SM{{\it Scripta Metall.}}
\def\ZAMM{{\it Z. Angew. Math. Mech.}}
\def\ZAMP{{\it Z. Angew. Math. Phys.}}
\def\ZVDI{{\it Z. Verein. Deut. Ing.}}

\def\salto#1#2{
[\mbox{\hspace{-#1em}}[#2]\mbox{\hspace{-#1em}}]}

\renewcommand\Affilfont{\itshape}
\setlength{\affilsep}{1em}
\renewcommand\Authsep{, }
\renewcommand\Authand{ and }
\renewcommand\Authands{ and }
\setcounter{Maxaffil}{2}

\title{Elastica catastrophe machine:
\\theory, design and experiments}

\author{Alessandro Cazzolli, Diego Misseroni, Francesco Dal Corso}
 \affil[]{DICAM, University of Trento, via~Mesiano~77, I-38123
Trento, Italy}

\date{}
\maketitle \footnotetext[0]{Corresponding author: Francesco Dal Corso fax:
+39 0461 282599; tel.: +39 0461 282522; web-site:
http://www.ing.unitn.it/$\sim$dalcorsf/; e-mail: francesco.dalcorso@unitn.it}

\begin{center}
\textit{Dedicated to our mentor and friend Davide Bigoni in honour of his 60$^{th}$ birthday,\\
for all of his invaluable  teaching throughout these years and for  many more to come
}
\end{center}

\date{}
\maketitle

\begin{abstract}
	The theory, the design and the experimental validation of a catastrophe machine based on a flexible element are addressed for the first time. A general  theoretical framework is developed by   extending that of the classical catastrophe machines made up of discrete elastic systems.  The new formulation, based on the nonlinear solution of the elastica, 
		is enhanced by considering  the concept of the   universal snap surface. Among the infinite set of  \emph{elastica catastrophe machines}, 
		two families are proposed and investigated to explicitly assess their features. The related catastrophe locus is disclosed in a large variety of shapes, very different from those generated by the classical counterpart. Substantial changes in the catastrophe locus properties, such as convexity and number of bifurcation points, are achievable by tuning the design parameters of the proposed machines towards the design of very efficient snapping devices. Experiments performed on the physical realization of the \emph{elastica catastrophe machine} fully validate the present theoretical approach. The developed model can find applications in mechanics  at different scales, for instance, in the design of new devices involving actuation or hysteresis loop mechanisms to achieve energy harvesting, locomotion, and wave mitigation.
\end{abstract}

\noindent{\it Keywords}: Nonlinear mechanics, snap mechanisms,
structural instability.

\section{Introduction}

Catastrophe theory is a well-established mathematical framework initiated by R. Thom \cite{thom}  for analyzing  complex systems exhibiting instability phenomena. 
	From its birth, concepts of this theory have been  exploited over the years in several fields to provide the interpretation of sudden large changes in the configuration as the result of a small variation in the boundary conditions. Owing to its  multidisciplinary application, catastrophe theory has found relevance  in the mechanics of fluids, solids, and structures \cite{cherepanov,grohpirrera,lengyel,ocarroll, qin,thompson,zeemaneuler},  but also in optics, physical chemistry, economics, biology and sociology \cite{arnold,gilmore,postonstew,thombook}.

\begin{figure}[H]
	\begin{center}
		\includegraphics[width=0.8\textwidth]{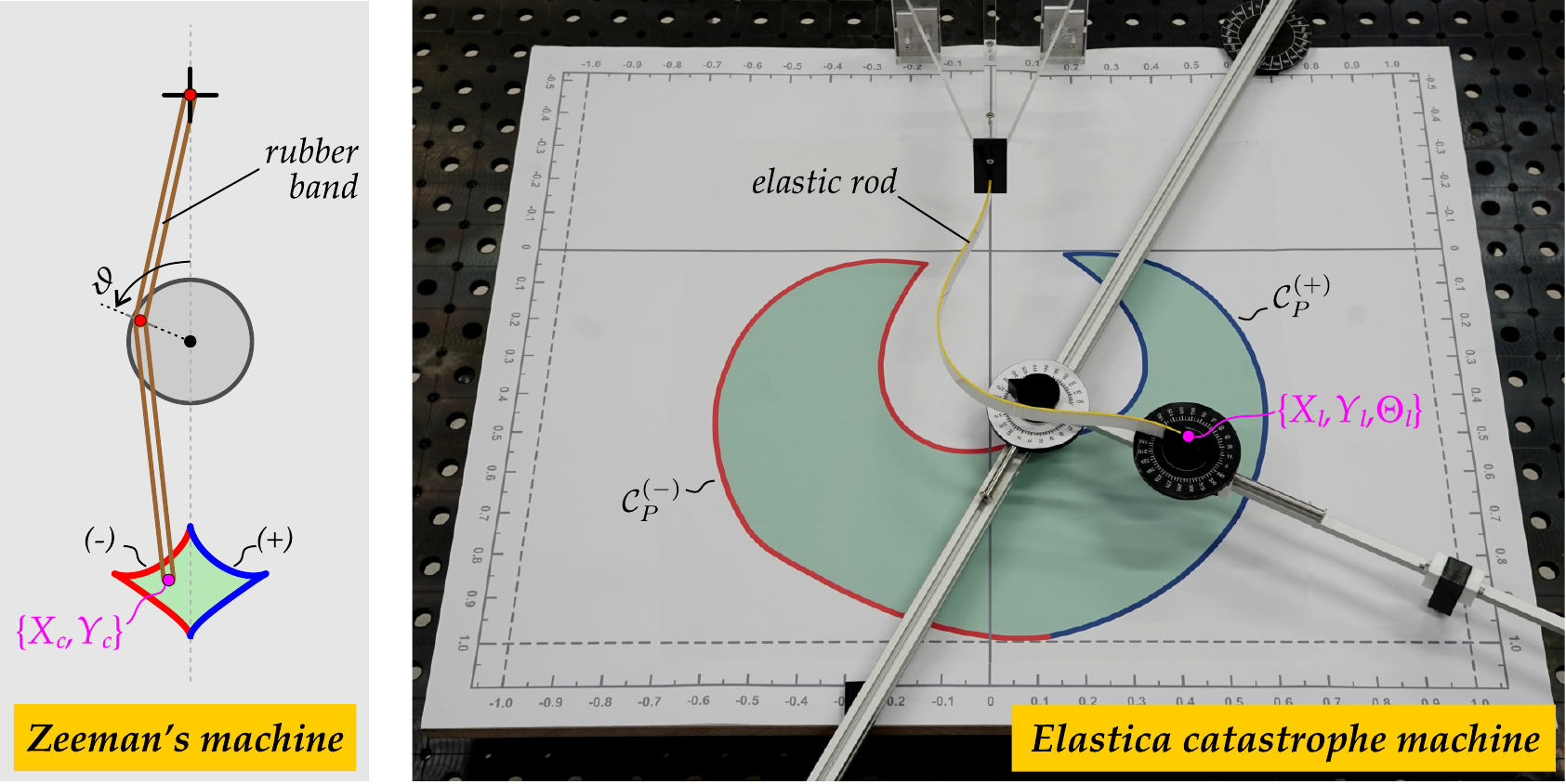}
	\end{center}
	\caption{\footnotesize{A sketch of the classical (discrete) catastrophe machine (left,  cf. Fig. 5.1 in \cite{postonstew}) and a photo of the  prototype realized for the proposed \emph{elastica catastrophe machine} (right). The respective catastrophe locus $\mathcal{C}_P$ is reported for both machines as the union of $\mathcal{C}_P^{(+)}$ (blue line) and $\mathcal{C}_P^{(-)}$ (red line). 
				Two stable equilibrium configurations exist for the elastic systems when the hand position, controlling the rubber's  end coordinates $X_c,Y_c$ (left) or the elastica's end coordinates $X_l,Y_l$, is within the bistable (green background) region enclosed by catastrophe locus. Differently,  the stable equilibrium configuration is unique when the hand position is located within the monostable region (non-green background) defined as outside of the closed curve defining the catastrophe locus. Crossing the catastrophe locus from inside may provide snapping of the system.
	}}
	\label{machines0intro}
\end{figure}

About fifty years ago,  E.C. Zeeman invented and realized a simple but intriguing 
mechanical device  \cite{zeeman0} to illustrate  for the first time concepts of catastrophe theory. The pioneering (planar)
two-spring system, also known as \lq Zeeman's catastrophe machine', can be easily home-built  by
fixing two elastic rubber bands  and a cardboard disk on a desktop through three drawing pins (Fig. \ref{machines0intro}, left).
More specifically, the two elastic bands are tied together through  a knot pinned on the cardboard disk. 
The other end of the first elastic band is pinned on the table while that of the second elastic band is held  by hand, controlling its position within the plane.
Lastly, in turn,  the disk is pinned to the desktop. 
The resulting system has \textit{two control parameters} (the hand position coordinates $X_c$ and $Y_c$) and \textit{one state variable} (the rotation angle $\vartheta$ of the cardboard disk). The number of equilibrium configurations for the system varies by changing the two control parameters (hand coordinates). 
	In  particular, the physical plane is split into two complementary regions separated by a  symmetric concave diamond-shaped curve (with four cusps): the monostable region outside the closed curve and the bistable region inside. These two regions  are respectively associated to  hand position providing either a unique or two different stable  equilibrium configurations (expressed by the state variable $\vartheta$).
	The separating closed curve is called the \textit{catastrophe locus}  because when crossed by the  hand position from inside to outside\footnote{
			Snapping  occurs only for the configuration inside the bistable region which loses stability when crossing the catastrophe locus. For the classical machine this is strictly related to the sign of the rotation angle $\vartheta$ and, similarly, for the presented elastica machine   to that of the curvature at the rod's ends.} provides the snapping of the system, as visual representation of the catastrophic behaviour.

Several modified versions of the Zeeman's catastrophe machine have been proposed with the purpose 
to display various concepts of catastrophe theory. Different two-spring \cite{hines} and three-spring \cite{yin} systems have been shown to possibly display more (than one) separated closed curves representing the \textit{catastrophe locus} by choosing  specific design parameters.
A different behaviour, the butterfly catastrophe, has been displayed when the elastic band pinned to the desktop is replaced by two identical elastic bands,  with their ends symmetrically pinned to the desktop \cite{woodcock}. The analysis of catastrophe locus has been also extended to discrete systems with  elastic hinges \cite{carricato2001,carricato2002}. 
Moreover, Zeeman's machine has also been used to show chaotic motion \cite{nagy}
and its principle has been exploited to motivate the electro-mechanical instabilities of a membrane under polar symmetric conditions \cite{lu}. However, the elastic response in all of these systems  has been considered to depend only on a finite number of degrees of freedom.

In this research line, the design of a catastrophe machine is extended for the first time to an elastic continuous element, namely the planar elastica, within the finite rotation regime.\footnote{The framework of catastrophe theory is found in the literature  to be only exploited for continuous systems in investigating their equilibrium configurations  as small perturbations of the undeformed one, as in the buckling problem for a pin ended rod under a lateral load \cite{zeemaneuler,zeemaneuler0} or for a stiffened plate \cite{hunt}. Differently, the catastrophe framework is here exploited for the whole set of equilibrium configurations, without any restriction on the amplitude of the related rotation field, being the analytical solutions of the Euler's elastica equation.} The increase of the number of degrees of freedom (from finite to infinite) together with the increase in the number of kinematic boundary conditions (from two to three) requires a more complex formulation in comparison with that considered for treating the classical discrete systems.

More specifically, considering as fixed the position of one end of the elastica, the three kinematic boundary conditions $X_l$, $Y_l$ (the two coordinates) and $\Theta_l$ (the rotation angle) at the other end are imposed through two control parameters. This relationship introduces a multiplicity issue for the configuration  associated with the same coordinates $X_l$, $Y_l$ of the final end (because of the sensitivity of the angular periodicity for the rotation angle) to be overcome for a proper representation of the catastrophe locus in the physical plane.
%

Furthermore, the analysis of  catastrophe loci for elastica based machines requires to consider a further space, the primary kinematical space, in addition to the two spaces usually considered in the analysis of classical machines, the control parameter  and the physical planes (no longer coincident here). It is shown that the catastrophe locus is provided by the projection in the physical plane of the intersection of the elastica machine set (defined by design parameters chosen for a specific machine) and the snap-back surfaces   (universal for elasticae with controlled ends \cite{cazzolli})  within the primary kinematical space. 

Among the infinite set of \emph{elastica catastrophe machines} (ECMs), two families  are proposed and thoroughly investigated through the developed theoretical formulation, fully confirmed by experiments performed on a physical model
(Fig. \ref{machines0intro}, right).

\begin{figure}[!htb]
	\begin{center}
		\includegraphics[width=0.8\textwidth]{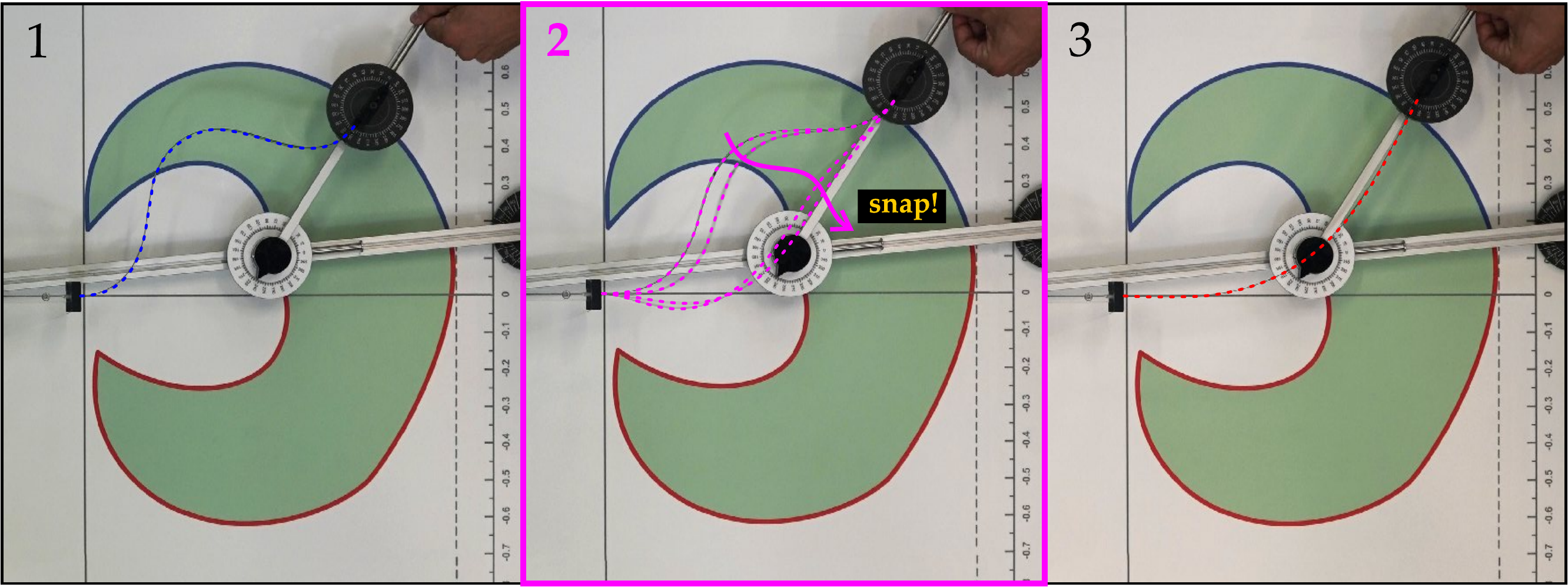}\\[2mm]
		\includegraphics[width=0.8\textwidth]{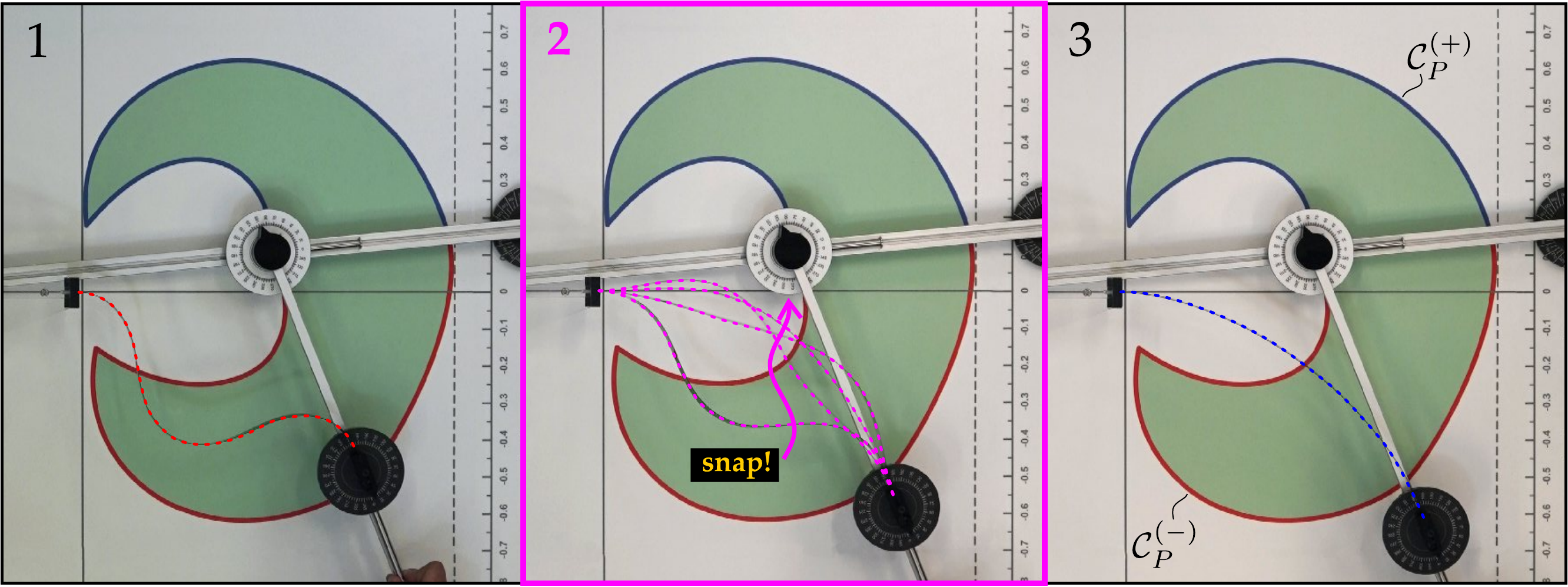}
	\end{center}
	\caption{\footnotesize{Evolution of the deformed configuration  for two different sequences in the rod's end position  controlled by hand.	
	Snapping occurs at crossing the catastrophe locus through the blue line $\mathcal{C}_P^{(+)}$ (upper part)/red line $\mathcal{C}_P^{(-)}$ (lower part) for the elastica having positive/negative curvature at its ends. Four snapshots taken during snapping are superimposed in the second column (deformed configurations highlighted with purple dashed lines).
	Experiments are performed using ECM-I (with $\kappa_R=0.5,\,\lambda_R=0.1,\,\upsilon=0$) with a carbon fiber rod by increasing the first control parameter $p_1$ (radial distance from the rotation point) at  fixed value of $p_2$ (the angle $\Theta_l$ at the moving end). Deformed configurations with positive/negative curvature at its ends are highlighted with blue/red dashed line.
			}}
	\label{fig_snapshot1}
\end{figure}

An example of snapping motion displayed by the realized physical model of the \emph{elastica catastrophe machine} (Fig.  \ref{machines0intro}, right) is illustrated in Fig.  \ref{fig_snapshot1}. Two sequences of deformed configurations are shown for two different evolutions of  the rod's final end position (controlled by hand). Both evolutions start from the  bistable (green) domain  (first column) and end to the monostable (white) domain (third column). Snapping occurs at crossing the catastrophe locus from inside to outside (second column highlighted in purple), as the elastic rod dynamically  reaches the reverted stable configuration.

A parametric analysis performed by varying design parameters shows that the  introduced families  define catastrophe loci in a large variety of shapes, very different from those realized with classical catastrophe machines. 
In contrast to the classical machines, it is shown that such sets may display unexpected geometrical properties. On one hand, the number of bifurcation points along the catastrophe locus can be different than four. On the other hand, the convexity measure \cite{convex} of catastrophe locus is found to change significantly, while that of classical machines (Fig. \ref{machines0intro}, left) is usually around 0.65.\footnote{Convex catastrophe loci can be found for force controlled discrete systems \cite{yin}. Nevertheless, convex catastrophe loci are not observed for classical catastrophe machines under displacement control.} In particular, the convex measure is found to possibly approach 1 with obtuse corners at the bifurcation points. This property facilitates reaching high-energy release snapping conditions, while these are difficult to attain in classical machines because associated with acute corner points.  

The combination of the variable number of bifurcation points and the approximately unit value for the convex measure paves the way to realize very efficient snapping devices. Therefore, in addition to the interesting mechanical and mathematical features with reference to catastrophe theory in combination with snapping mechanisms \cite{armanini,bigoniCISM,camescasse,chen2008,chen2012, fargette,sano1, sano2, schioler}, the proposed model may find application in the design of cycle mechanisms for actuation and dissipation devices towards energy harvesting, locomotion and wave mitigation \cite{bertoldi,cohen,fraternali,harne,koch, raney,yang}.

\section{Equilibrium configurations for the elastica and the universal snap surface}

The equilibrium configurations and the concept of universal snap surface are recalled for the \emph{inextensible planar elastica} of length $l$ and lying within the $X-Y$  plane, which models the flexible element composing the \emph{elastica catastrophe machine}.
Considering the flexible element (the rod) kinematically constrained at its two ends, the following  six boundary conditions are imposed
\beq\label{bcbcbc}
\barr{ccc}
X(s=0)=X_0,\qquad Y(s=0)=Y_0,\qquad \Theta(s=0)=\Theta_0,\\[4mm]
X(s=l)=X_l,\qquad Y(s=l)=Y_l,\qquad \Theta(s=l)=\Theta_l,
\earr
\eeq
where $s\in [0,l]$ denotes arc length along the rod, $X$ and $Y$ the Cartesian coordinates and $\Theta$ the anticlockwise rotation evaluated with respect to the $X$ axis.  The inextensibility of the elastic rod constrains the distance $d$ between its two ends to satisfy the following kinematic compatibility condition
\begin{equation}
\label{distanced}
d(X_0, Y_0, X_l, Y_l)=\sqrt{(X_l-X_0)^2+(Y_l-Y_0)^2}\leq l.
\end{equation}

The inextensibility assumption also introduces the dependence of the coordinate fields $X(s)$ and $Y(s)$ on the rotation field $\Theta(s)$ through the following differential relations
\begin{equation}
\label{inext}
X'(s)=\cos{\Theta(s)}, \qquad Y'(s)=\sin{\Theta(s)},
\end{equation}
where the symbol $\,'\,$ denotes the derivative with respect to the curvilinear coordinate $s$.

Given the six boundary conditions (\ref{bcbcbc}),  the deformed configuration of the elastic rod at equilibrium is described by
\begin{equation}\label{rotasol}
\left\{
\begin{array}{lll}
X(s)=\ds X_0+\frac{X_l-X_0}{d}\mathsf{C}(s)l
-\frac{Y_l-Y_0}{d}\mathsf{D}(s)
l,\\[4mm]
Y(s)=\ds Y_0+\frac{X_l-X_0}{d}\mathsf{D}(s)l
+\frac{Y_l-Y_0}{d}\mathsf{C}(s)
l,\\[4mm]
\Theta(s)= \ds\arctan{\left[\frac{Y_l-Y_0}{X_l-X_0}\right]}+\beta+2\zeta(s),
\end{array}
\right.
\end{equation}
where  $\beta$ is related to the inclination of the reaction force at the ends, measured as anti-clockwise angle
with respect to the straight line connecting the two clamps, while 
\beq
\mathsf{C}(s)=\mathsf{A}(s)\cos{\beta}+\mathsf{B}(s)\sin{\beta},\qquad 
\mathsf{D}(s)=\mathsf{A}(s)\sin{\beta}-\mathsf{B}(s)\cos{\beta}.
\eeq

In the case when the number $m$ of inflection points along the rod is null, the three functions $\zeta(s)$, $\mathsf{A}(s)$, and $\mathsf{B}(s)$ are given by
\begin{equation}\label{rotasol3}
\left\{
\begin{array}{lll}
\zeta(s)= 
\,\textup{am}\left(\dfrac{s}{l}\left(F(\zeta_{l},\,\xi)-F(\zeta_{0},\,\xi)\right)+F(\zeta_{0},\,\xi),\,\xi\right), \\[4mm]
\mathsf{A}(s)= 
\ds\frac{2}{\xi^2}\frac{\mathscr{E}\left(\dfrac{s}{l}\left(F(\zeta_{l},\,\xi)-F(\zeta_{0},\,\xi)\right)
	+F(\zeta_{0},\,\xi),\,\xi\right)+\mathscr{E}\left(F(\zeta_{0},\,\xi),\,\xi\right)}{F(\zeta_{l},\,\xi)-F(\zeta_{0},\,\xi)}-\frac{2-\xi^2}{\xi^2}\frac{s}{l},\\
\mathsf{B}(s)= 
\ds\frac{2}{\xi^2}\frac{\textup{dn}\left(\dfrac{s}{l}\left(F(\zeta_{l},\,\xi)-F(\zeta_{0},\,\xi)\right)+F(\zeta_{0},\,\xi),\,\xi\right)
	-\textup{dn}\left(F(\zeta_{0},\,\xi),\,\xi\right)}{F(\zeta_{l},\,\xi)-F(\zeta_{0},\,\xi)}.
\end{array}
\right.
\end{equation}
Differently, when at least one inflection point is present ($m\neq0$),
\begin{equation}\label{rotasol4}
\left\{
\begin{array}{lll}
\zeta(s)= 
\arcsin{\left[\eta\,\textup{sn}\left(\dfrac{s}{l}\left(F(\omega_{l},\,\eta)-F(\omega_0,\,\eta)\right)+F(\omega_0,\,\eta),\,\eta\right)\right]}, \\[4mm]
\mathsf{A}(s)= 
\ds
\ds2\frac{\mathscr{E}\left(\dfrac{s}{l}\left(F(\omega_{l},\,\eta)-F(\omega_0,\,\eta)\right)+F(\omega_0,\,\eta),\,\eta\right)
	-\mathscr{E}\left(F(\omega_0,\,\eta),\,\eta\right)}{F(\omega_{l},\,\eta)-F(\omega_0,\,\eta)}
-\frac{s}{l},\\
\mathsf{B}(s)= 
\ds 2\,\eta\,\frac{\textup{cn}\left(\dfrac{s}{l}\left(F(\omega_{l},\,\eta)-F(\omega_0,\,\eta)\right)+F(\omega_0,\,\eta),\,\eta\right)
	-\textup{cn}\left(F(\omega_0,\,\eta),\,\eta\right)}{F(\omega_{l},\,\eta)-F(\omega_0,\,\eta)}.
\end{array}
\right.
\end{equation}

	In the aforementioned equations $F$ is \textit{Jacobi's incomplete elliptic integral of the first kind}, 
	$\mathscr{E}$ \textit{Jacobi's epsilon function}, $E$ \textit{Jacobi's incomplete elliptic integral of
		the second kind}, \lq $\textup{sn}$'  \textit{Jacobi's sine amplitude function},  \lq $\textup{cn}$'  \textit{Jacobi's cosine amplitude function}, \lq $\textup{dn}$'  \textit{Jacobi's elliptic function}, and \lq $\textup{am}$'   \textit{Jacobi's amplitude function},
\begin{equation}
\barr{ccc}
F\left(\varphi,\,k\right)=\ds\int_{0}^{\varphi}\frac{\textup{d}\phi}{\sqrt{1-k^2\sin^2{\phi}}},
\,
E\left(\varphi,\,k\right)=\int_{0}^{\varphi}\sqrt{1-k^2\sin^2{\phi}}\,\textup{d}\phi,
\, \mathscr{E}\left(\varphi,\,k\right)=E(\textup{am}(\varphi,\,k),\,k),
\\[6mm]
\textup{sn}(u,\,k)=\sin{\left(\textup{am}(u,\,k)\right)},\qquad
\textup{cn}(u,\,k)=\cos{\left(\textup{am}(u,\,k)\right)},\\[4mm]
\varphi=\textup{am}\bigg(F\left(\varphi,\,k\right),\,k\bigg),
\qquad
\textup{dn}(u,\,k)=\sqrt{1-k^2\,\textup{sn}^2(u,\,k)}.
\earr
\end{equation}

Moreover, the parameters $\zeta_0$, $\zeta_{l}$, $\eta$, $\omega_0$, and
$\omega_{l}$ appearing in   eqns (\ref{rotasol3}) and (\ref{rotasol4}) are given by
\begin{equation}\label{carorevisore}
\begin{array}{ccc}
\zeta_0=\dfrac{\Theta_0+\beta}{2}-\ds\frac{1}{2}\arctan{\left[\frac{Y_l-Y_0}{X_l-X_0}\right]},\qquad
\zeta_{l}=\dfrac{\Theta_l+\beta}{2}-\frac{1}{2}\arctan{\left[\frac{Y_l-Y_0}{X_l-X_0}\right]},\qquad
\eta=\left|\sin{\hat{\zeta}}\right|,
\\[6mm]
\omega_0=\arcsin{\left(\dfrac{\sin{\zeta_0}}{\eta}\right)},\qquad
\omega_{l}=(-1)^m\,\arcsin{\left(\dfrac{\sin{\zeta_l}}{\eta}\right)}+(-1)^j\,m\pi,
\end{array}
\end{equation}
with $\hat{\zeta}=\zeta(\hat{s})$, $\hat{s}$ being the smallest curvilinear coordinate $s$ corresponding to an inflection point, $\Theta'(\hat{s})=0$, and the parameter $j$ related to the sign of curvature at $s=0$ 
(corresponding to $j=0$ if $\Theta'(s=0)>0$, and $j=1$ if $\Theta'(s=0)<0$), while $\xi$ is a parameter restricted to
\beq
\xi\in\left[0,\,\sqrt{\ds\frac{2}{1-\ds\min_{s\in[0,l]}\left\{\cos 2\zeta(s)\right\}}}\right].
\eeq

\begin{figure}[H]
	\begin{center}
		\includegraphics[width=0.8\textwidth]{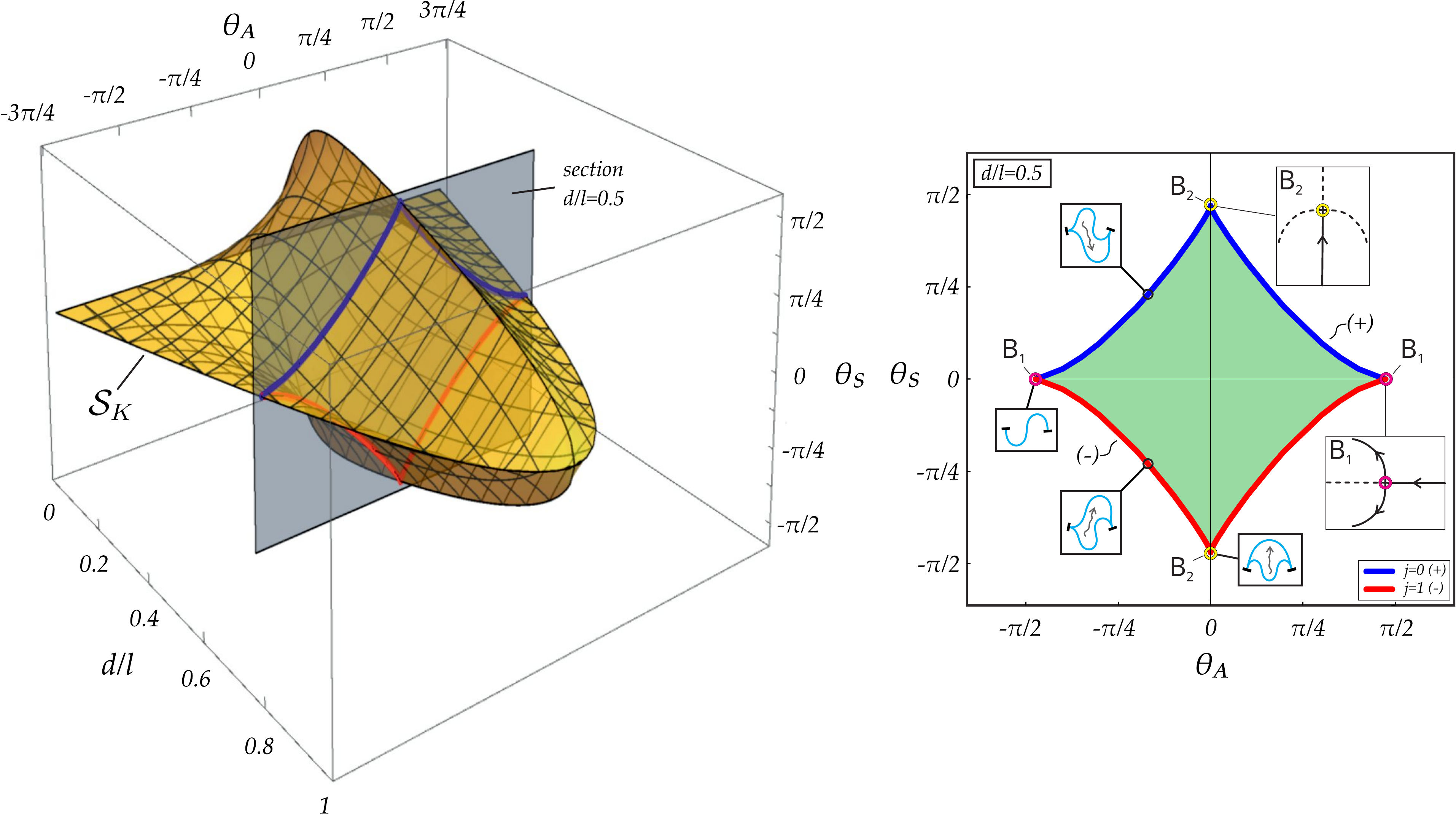}
	\end{center}
	\caption{\footnotesize{(Left) Universal snap surface $S_k$ (type 1) within the space of the primary kinematical quantities $\{d,\theta_A,\theta_S\}$ and its section with the plane $d=0.5l$. (Right) Snap surface section for $d=0.5 l$  as thick closed curve  within the plane $\theta_A-\theta_S$  composed by blue $(+)$ and red  $(-)$ parts. The blue (red) part refers to snapping configurations with positive  $j=0$ (negative $j=1$) curvature at both ends. The bistable (monostable) domain is reported as the green (white) region inside (outside) the closed curve. The deformed shapes of the elastica before and after snapping  are reported as insets for some critical condition along the closed curve are reported.  The two cusps $\mathsf{B}_1$ ($\mathsf{B}_2$) are supercritical (subcritical) pitchfork bifurcations, and are associated with a non-snapping (a snapping) configuration. 
	}}
	\label{section_sk}
\end{figure}

The position fields $X(s)$, $Y(s)$, and $\Theta(s)$, eqn (\ref{rotasol}), define the  configuration taken by the elastica when constrained by the two ends. In particular, the equilibrium configuration (in general non-unique) can be characterized once the two unknown parameters ($\xi$ and $\beta$ for $m=0$,  $\eta$ and $\beta$ for $m\neq0$) are evaluated for a given set of kinematical boundary conditions, eqn (\ref{bcbcbc}). The pair(s) of these parameters can be obtained by solving the following nonlinear system,
\begin{equation}
\label{sistPOSnoflex}
\left\{
\begin{aligned}
& \mathsf{C}(l) = \frac{d}{l},\\
& \mathsf{D}(l)= 0.
\end{aligned}
\right.
\end{equation}
Towards the stability analysis of a specific equilibrium configuration related to the six boundary conditions $X_0$, $X_l$, $Y_0$, $Y_l$, $\Theta_0$, $\Theta_l$, it is instrumental to refer to the following three primary kinematical quantities: the  distance $d$, eqn (\ref{distanced}), and the angles $\theta_A$ and $\theta_S$, respectively defined as the  antisymmetric and symmetric parts of the imposed end rotations,
\begin{equation}
\label{tt}
\begin{split}
\theta_A=\frac{\Theta_l+\Theta_0}{2}-\arctan{\left[\frac{Y_l-Y_0}{X_l-X_0}\right]}, \qquad
\theta_S=\frac{\Theta_l-\Theta_0}{2}.
\end{split}
\end{equation}
In particular, the triads $\{d,\theta_A,\theta_S\}$   can be related to a unique or two different stable configurations through a function $S_K (d,\,\theta_A,\,\theta_S)$ as \cite{cazzolli}
\beq
\label{bimono}
\begin{array}{lll}
	S_K (d,\,\theta_A,\,\theta_S)>0\qquad\Leftrightarrow \qquad\mbox{monostable domain: one stable configuration},\\[4mm]
	S_K (d,\,\theta_A,\,\theta_S)<0\qquad\Leftrightarrow \qquad\mbox{bistable domain: two stable configurations}.
\end{array}
\eeq
\paragraph{Universal snap surface.}
	The transition between the bistable and monostable domains (\ref{bimono}) occurs for the set of critical conditions of snap-back for one of the two stable configurations, differing by the sign of curvature at the two ends. Such a condition can be represented through the concept of \emph{universal snap surface} (restricted here to type 1 only \cite{cazzolli}),  which can be expressed in the following implicit form
\beq\label{Sk}
S_K (d,\,\theta_A,\,\theta_S)=0\qquad\Leftrightarrow \qquad\mbox{one stable and one critical configuration at snap},
\eeq
The equation (\ref{Sk}) defines a closed surface within the space of the primary kinematical quantities $\{d,\theta_A,\theta_S\}$, with two planes of symmetry  defined by $\theta_A=0$ and $\theta_S=0$ (Fig. \ref{section_sk}, left).\footnote{It is noted that a type 1 snapping configuration is always related to an elastica with two inflection points, $m=2$, which snaps towards another elastica with two inflection points. Therefore, each configuration at snapping displays the same curvature the sign at both ends, changing sign from just before to just after the snap mechanism.} 
	The intersection of the surface $S_K$ with its two symmetry planes provides two closed curves representing the whole set of pitchfork bifurcation points. More specifically, the pitchfork bifurcation points are distinguished as \emph{supercritical} or \emph{subcritical}, the former corresponding to the intersection curve with $\theta_S=0$  and the latter with $\theta_A=0$. Therefore, the generic planar section of $S_K$ at fixed values of  $d$ has shape and  physical meaning definitely similar to those of the catastrophe locus of the classical Zeeman machine (see Fig. \ref{machines0intro} left) having two canonical and two dual cusps, see \cite{cazzolli} and \cite{postonstew}.
A critical configuration with a certain sign of curvature at the two ends is characterized by symmetric angle $\theta_S$ with the same sign. Due to the symmetry properties described above, the implicit function $S_K (d,\,\theta_A,\,\theta_S)$ can be described through two single value  functions $\theta_S^{sb\,(+)}$ and $\theta_S^{sb\,(-)}$  of the  two primary kinematical quantities $d$ and $\theta_A$,
\beq\label{thetaSsnap}
\theta_S^{sb\,(+)}=\theta_S^{sb\,(+)}(d,\theta_A),\qquad
\theta_S^{sb\,(-)}=\theta_S^{sb\,(-)}(d,\theta_A),
\eeq
where the sign enclosed by the superscript parentheses is related to the sign of curvature at the two ends before snapping, related to the parameter $j$ in eqn (\ref{carorevisore}). This sign is also coincident with that of the symmetric angle of the snapping configuration. Furthermore, symmetry properties lead to the following conditions
\beq
\theta_S^{sb\,(+)}(d,\theta_A)=\theta_S^{sb\,(+)}(d,-\theta_A)=
-\theta_S^{sb\,(-)}(d,\theta_A)=-\theta_S^{sb\,(-)}(d,-\theta_A).
\eeq

\section{Theoretical framework for \emph{elastica catastrophe machines}}

The aim of this section is to develop the theoretical framework for the realization of the \emph{elastica catastrophe machine}. 	For the sake of simplicity,  the initial coordinate of the elastic rod, $s=0$, is considered fixed and taken as the origin of the  reference system $X-Y$
	and with tangent parallel to the $X$-axis, so that
\beq\label{assumption}
X_0=Y_0=\Theta_0=0.
\eeq
It follows that the three primary kinematical quantities, eqn (\ref{tt}) can be expressed as functions (overtilde symbol) of the position at the final coordinate ($s=l$) only, 
\begin{equation}\label{kintilde}
d=\widetilde{d}(X_l,Y_l),\qquad
\theta_A=\widetilde{\theta}_A(X_l,Y_l,\Theta_l),\qquad
\theta_S=\widetilde{\theta}_S(\Theta_l),
\end{equation}
which simplify as\footnote{Details about overcoming  periodicity issues inherent to the trigonometric function arctan are reported in Section \ref{appA_1} of the Supplementary Material.}

\begin{equation}
\label{tt2}
\begin{split}
\widetilde{d}(X_l,Y_l)=\sqrt{X_l^2+Y_l^2},\qquad
\widetilde{\theta}_A(X_l,Y_l,\Theta_l)=\frac{\Theta_l}{2}-\arctan{\left[\frac{Y_l}{X_l}\right]}, \qquad
\widetilde{\theta}_S(\Theta_l)=\frac{\Theta_l}{2},
\end{split}
\end{equation}
Relations (\ref{tt2}) can be inverted to provide the position at the final coordinate ($s=l$) as a function (hat symbol) of the primary kinematical quantities
\beq\label{icshat}
\widehat{X}_l(d,\theta_A,\theta_S)= d \cos(\theta_S-\theta_A),
\qquad
\widehat{Y}_l(d,\theta_A,\theta_S)= d \sin(\theta_S-\theta_A),
\qquad
\widehat{\Theta}_l(\theta_S)= 2 \theta_S.
\eeq
The definition of a (catastrophe) machine leads to the introduction of control and design parameters, respectively collected in the two vectors  $\textbf{p}=\{p_1,...,p_M\}$ and $\textbf{q}=\{q_1,...,q_N\}$ (with $M,N\in\mathbb{N}$). In particular, $\textbf{p}$ is the fundamental vector collecting the degrees of freeedom of the considered machine. Thus, the position of the rod at the final curvilinear coordinate ($s=l$) can be also described as functions (overbar symbol) of such parameters as
\beq\label{positionbar}
X_l=\overline{X}_l(\textbf{p}, \textbf{q}),\qquad
Y_l=\overline{Y}_l(\textbf{p}, \textbf{q}),\qquad
\Theta_l=\overline{\Theta}_l(\textbf{p}, \textbf{q}),
\eeq
and similarly, considering eqns (\ref{tt2}) and (\ref{positionbar}), the  three primary kinematical quantities $d,\theta_A$ and $\theta_S$ as
\beq\label{eqKC}
d=\overline{d}(\textbf{p}, \textbf{q}),\qquad
{\theta}_A=\overline{\theta}_A(\textbf{p}, \textbf{q}),\qquad
{\theta}_S=\overline{\theta}_S(\textbf{p}, \textbf{q}).
\eeq
Although both the control and design parameters affect the elastica configuration, a distinction is made  being the former varied at fixed values of the latter. 

In the following, towards the geometrical representation of the \emph{catastrophe locus} (namely, the  critical  conditions providing snapping for the elastica) within the physical plane $X-Y$, the number of control parameters is taken as $M=2$, so that $\textbf{p}=\{p_1,p_2\}$. 

Finally, it is assumed that the relations (\ref{positionbar}) and (\ref{eqKC}) can be inverted, thus obtaining
\beq\label{ptilda}
p_j=\widetilde{p}_j(X_l,Y_l,\Theta_l, \textbf{q})\qquad \qquad j=1,2,
\eeq
and
\beq\label{phat}
p_j=\widehat{p}_j(d,\theta_A,\theta_S, \textbf{q})\qquad \qquad j=1,2,
\eeq
respectively.
A generic configuration of the elastica can be therefore represented by the three equivalent parametrisations of the boundary conditions, namely i) by the two control parameters $\{p_1,p_2\}$, ii) by the three coordinates of the rod's final end $\{X_l,Y_l,\Theta_l\}$ or iii) by the three primary kinematical quantities $\{d,\theta_A,\theta_S\}$. This discrepancy in the number of the required parameters suggests that one of the coordinates of the triads $\{X_l,Y_l,\Theta_l\}$ or $\{d,\theta_A,\theta_S\}$ might be expressed as a function of the remaining two. Section \ref{appA_1} of the Supplementary Material is devoted to the development of such statement.

\subsection{Three spaces for representing the catastrophe locus}
In light of the above, the complete understanding of the principles underlying the present catastrophe machine requires to consider the projection of the controlled end's configuration within the three different spaces,
\begin{itemize}
	\item[$C$:]
	the control parameter plane $p_1-p_2$;
	\item[$K$:] the primary kinematical quantities space $d-\theta_A-\theta_S$;
	\item[$P$:] the physical space $X_l-Y_l-\Theta_l$, where the rotational coordinate $\Theta_l$ (possibly even more than one) is condensed to the physical plane $X_l-Y_l$. 
\end{itemize}
The need of these three different representations and the (unavoidable) projection of the rotational coordinate $\Theta_l$ to the physical plane $X_l-Y_l$ are the new constituents of the \emph{elastica catastrophe machine} with respect to the classical one \cite{zeeman0,zeeman}, where the control plane coincides with the physical one and the kinematical space is not needed.
Furthermore, differently from the classical catastrophe machine, the values of the control parameters $p_1$ and $p_2$, kinematical quantities $d$, $\theta_A$, and $\theta_S$, and the end's coordinates $X_l$, $Y_l$, and $\Theta_l$ are here  restricted by the inextensibility constraint, eqn (\ref{distanced}), so that their variation is limited to the \lq inextensibility set' $\mathcal{I}$, which is defined in the three different spaces  as
\beq
\begin{array}{lll}
	\mathcal{I}_C:=\{\textbf{p}\in \mathbb{R}^2 \big|\,0\leq \overline{d}(\textbf{p},\textbf{q})\leq l\},
	\\[2mm]
	\mathcal{I}_K:=\{\{d,\,\theta_A,\,\theta_S\}\in \mathbb{R}^3 \big|\,0\leq d\leq l\},
	\\[2mm]
	\mathcal{I}_P:=\{\{X_l,\,Y_l\}\in \mathbb{R}^2 \big|\,0\leq \widetilde{d}(X_l,\,Y_l)\leq l\}.
\end{array}
\eeq
In order to minimize the presence of self-intersecting configurations\footnote{The considered limitation for the values of the angles $\theta_A$ and $\theta_S$ also provides that the machine set does not display type 2 snapping mechanisms  \cite{cazzolli}.} for the elastica,  the symmetric $\theta_S$ and antisymmetric $\theta_A$ angles are considered to be restricted by
\beq\label{limitazionesymasym}
\left\{\left|\theta_A\right|,\left|\theta_S\right|\right\}<\pi,
\eeq
so that the variables within the three spaces are also limited  to the \lq machine set' $\mathcal{M}$, 
\beq
\begin{array}{lll}
	\mathcal{M}_C := \left\{\textbf{p}\in \mathbb{R}^2 \big|\,\left\{\left|\overline{\theta}_A(\textbf{p},\textbf{q})\right|,\left|\overline{\theta}_S(\textbf{p},\textbf{q})\right|\right\}<\pi\right\},
	\\[2mm]
	\mathcal{M}_K:=\left\{\{d,\,\theta_A,\,\theta_S\}\in \mathbb{R}^3 \big|\, \{d=\overline{d}(\textbf{p},\textbf{q}),\,\theta_A=\overline{\theta}_A(\textbf{p},\textbf{q}),\,\theta_S=\overline{\theta}_S(\textbf{p},\textbf{q})\}, \textbf{p}\in \mathcal{M}_C \right\},
	\\[2mm]
	\mathcal{M}_P:=\left\{\{X_l,\,Y_l\}\in  \mathbb{R}^2 \big|\, \{X_l=\overline{X}_l(\textbf{p},\textbf{q}),\,Y_l=\overline{Y}_l(\textbf{p},\textbf{q})\}, \textbf{p}\in \mathcal{M}_C \right\}.
\end{array}
\eeq
The intersection of the inextensibility $\mathcal{I}$ and machine $\mathcal{M}$ sets provides the \lq elastica machine set' $\mathcal{E}$, defining the configurations that can be attained by the designed \emph{elastica catastrophe machine},
\beq
\mathcal{E}_J:=\mathcal{M}_J \cap \mathcal{I}_J,\qquad J=C,K,P.
\eeq
The two single-valued  functions $\theta_S^{sb\,(+)}(d,\theta_A)$ and $\theta_S^{sb\,(-)}(d,\theta_A)$, eqn (\ref{thetaSsnap}),
introduced in the previous section as the collection of critical snap-back (type 1 \cite{cazzolli}) conditions   for positive and negative sign of ends' curvature configurations,  define respectively the \lq snap-back subsets' $\mathcal{S}_K^{(+)}$ and  $\mathcal{S}_K^{(-)}$ within the $d-\theta_A-\theta_S$ space
\beq
\begin{array}{lll}
	\mathcal{S}_K^{(+)}:=\left\{\left\{d,\,\theta_A,\,\theta_S\right\}\in \mathcal{I}_K \big|\, \theta_S=\theta_S^{sb\,(+)}(d,\theta_A)\right\},\\[5mm]
	\mathcal{S}_K^{(-)}:=\left\{\left\{d,\,\theta_A,\,\theta_S\right\}\in \mathcal{I}_K \big|\, \theta_S=\theta_S^{sb\,(-)}(d,\theta_A)\right\}.
\end{array}
\eeq
The union of the two \lq snap-back subsets' $\mathcal{S}_K^{(+)}$ and  $\mathcal{S}_K^{(-)}$ provides the \lq snap-back set' $\mathcal{S}_K$
\beq
\mathcal{S}_K=\mathcal{S}_K^{(+)}\cup \mathcal{S}_K^{(-)},
\eeq
splitting the $d-\theta_A-\theta_S$ space into two regions, the \lq bistable set' $\mathcal{B}_K$ collecting kinematical quantities for which two stable solutions exist
\beq
\mathcal{B}_K:=\left\{\{d,\,\theta_A,\,\theta_S\}\in \mathcal{I}_K \big|\, S_K \left(d,\theta_A,\theta_S\right)<0\right\},
\eeq
and the \lq monostable set' $\mathcal{U}_K$ collecting kinematical quantities for which only one stable solution exists
\beq
\mathcal{U}_K:=\left\{\{d,\,\theta_A,\,\theta_S\}\in \mathcal{I}_K \big|\, S_K \left(d,\theta_A,\theta_S\right)>0\right\}.
\eeq
The snap-back set $\mathcal{S}_K$ is independent from the design parameters and has only a representation within the primary kinematical space. Its intersection with the  \lq elastica machine set' $\mathcal{E}_K$  provides the critical kinematical quantities $d^{\mathcal{C}},\,\theta_A^{\mathcal{C}}$, and $\theta_S^{\mathcal{C}}$ associated with the designed elastica machine and collected in the  \lq catastrophe set'\footnote{It is worth mentioning that the present nomenclature differs from that used by some authors \cite{carricato2002,hines} defining the projection $\mathcal{C}_C$ of the catastrophe set in the control (force) plane as bifurcation set and the snap-back set $\mathcal{S}_K$ as catastrophe set.} $\mathcal{C}_K$
\beq
\mathcal{C}_K:=\mathcal{S}_K\cap \mathcal{E}_K= \left\{\{d^{\mathcal{C}},\,\theta_A^{\mathcal{C}},\,\theta_S^{\mathcal{C}}\}\in \mathcal{I}_K \big|\, S_K \left(d^{\mathcal{C}}=\overline{d}(\textbf{p},\textbf{q}),\,\theta_A^{\mathcal{C}}=\overline{\theta}_A(\textbf{p},\textbf{q}),\,\theta_S^{\mathcal{C}}=\overline{\theta}_S(\textbf{p},\textbf{q})\right)=0\right\}.
\eeq 
Considering eqns (\ref{positionbar}) and (\ref{pscappello}), the \lq catastrophe set' (or, equivalently, the \textit{catastrophe locus}) $\mathcal{C}_K$ can be also projected within the control and physical\footnote{It is noted that the  \lq catastrophe set' 
	$\mathcal{C}_P$ is a curve within the physical plane $X_l-Y_l$, obtained as the  projection  of the \lq catastrophe set' $\mathcal{C}_P^{3D}$ collecting the critical rotation  angle $\Theta_l^{\mathcal{C}}$ as third physical  coordinate 
	\beq
	\mathcal{C}_P^{3D}:= \{\{X_l^{\mathcal{C}},\,Y_l^{\mathcal{C}},\,\Theta_l^{\mathcal{C}}\}\in \mathcal{I}_P \big|\, X_l=\overline{X}_l(\textbf{p}^{\mathcal{C}},\textbf{q}),\,Y_l=\overline{Y}_l(\textbf{p}^{\mathcal{C}},\textbf{q}),\,\Theta_l=\overline{\Theta}_l(\textbf{p}^{\mathcal{C}},\textbf{q})\} .
	\eeq} planes
\beq
\begin{array}{lll}
	\mathcal{C}_C:= \{\textbf{p}^{\mathcal{C}}\in \mathcal{I}_C \big|
	\textbf{p}^{\mathcal{C}}=\widehat{\textbf{p}}\left(\theta_A^{\mathcal{C}},\,\theta_S^{\mathcal{C}}\right)\}, \qquad
	\mathcal{C}_P:= \{\{X_l^{\mathcal{C}},\,Y_l^{\mathcal{C}}\}\in \mathcal{I}_P \big|\, X_l=\overline{X}_l(\textbf{p}^{\mathcal{C}},\textbf{q}),\,Y_l=\overline{Y}_l(\textbf{p}^{\mathcal{C}},\textbf{q}) \}.
\end{array}
\eeq
Similarly to the  \lq snap-back set' $\mathcal{S}_K$, the  \lq catastrophe sets' $\mathcal{C}_J$ are given by the union of  the two  \lq catastrophe subsets' $\mathcal{C}^{(+)}_J$ and $\mathcal{C}^{(-)}_J$ ($J=C,K,P$), being the sign referred to that of the symmetric angle/ends curvature for which the equilibrium configuration snaps.  Due to the  nonlinearities involved, the catastrophe sets can be evaluated only numerically. The algorithm used for the numerical evaluation of the catastrophe set is presented in Section \ref{appA_3} of the Supplementary Material.

\subsection{\lq Effectiveness' of the \emph{elastica catastrophe machine}}
Following the principles of the classical catastrophe machine, the \lq effective' \emph{elastica catastrophe machine} should repetitively display snapping mechanisms along specific equilibrium paths. This property corresponds to the \emph{hysteretic} behaviour typical of nonlinear elastic structures characterized by cusp catastrophes when subject to cyclic variations in their control parameters \cite{postonstew}. Therefore, the design of an \lq effective' \emph{elastica catastrophe machine} is guided by tuning the design parameters $\bq$ towards the morphogenesis of an \lq effective' catastrophe set $\mathcal{C}_P$ displaying hysteresis. In particular, this set defines a closed curve in the physical plane which is composed of both the \lq catastrophe subsets' $\mathcal{C}^{(+)}_P$ and $\mathcal{C}^{(-)}_P$ joined together, allowing snapping for both signs of symmetric angle/ends curvature.

\begin{figure}[H]
	\begin{center}
		\includegraphics[width=0.85\textwidth]{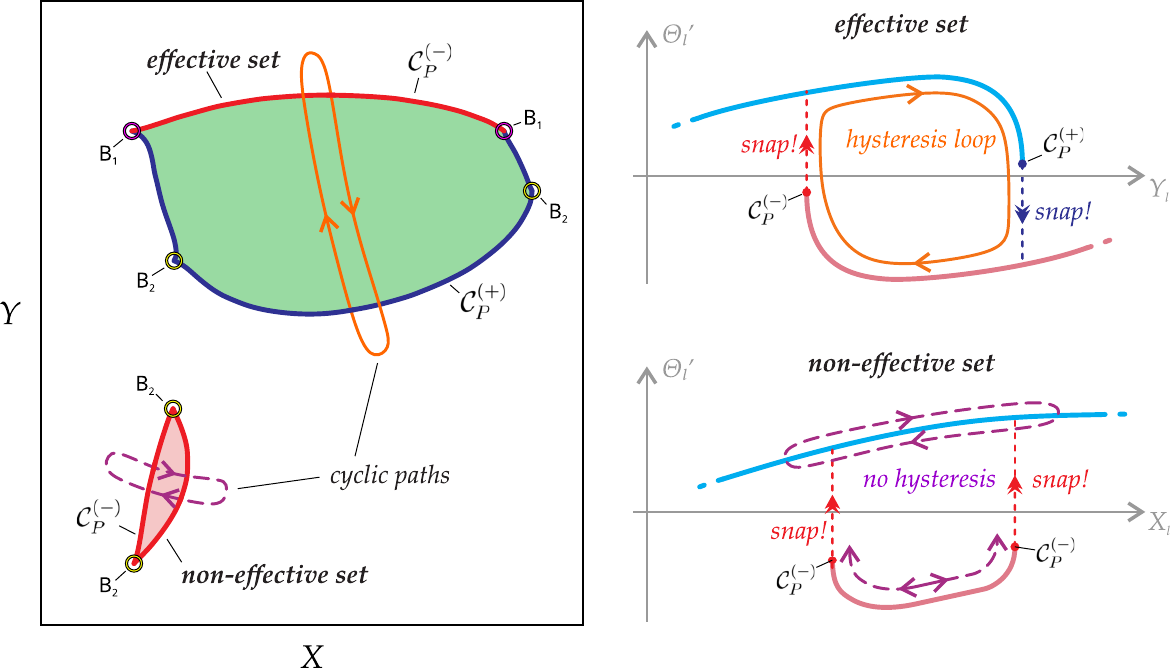}
	\end{center}
	\caption{\footnotesize{(Left)  Sketch of effective and non-effective catastrophe sets within the physical plane, the former composed of the two subsets $\mathcal{C}^{(+)}_P$ (blue curve) and $\mathcal{C}^{(-)}_P$ (red curve) while the latter of a unique closed subset $\mathcal{C}^{(-)}_P$ (red closed curve). Cyclic  paths crossing the two sets are drawn as  orange and purple closed loops, respectively. Green areas represent the set of coordinates $\{X_l,Y_l\}$ of the rod's end corresponding to bistability of the equilibrium. Sharp corner points $\mathsf{B}_1$ (magenta) and $\mathsf{B}_2$ (yellow) denote supercritical and subcritical bifurcations for the elastica with controlled ends, respectively. (Right) Sketch of the structural response in terms of the end's curvature $\Theta_l'$ versus the evolution of the  end's coordinate $X_l$ ($Y_l$) along the orange/continuous (purple/dashed) cyclic path and providing hysteretic (non-hysteretic) behaviour.
	}}
	\label{fig_hysteresis}
\end{figure}

The hysteretic (non-hysteretic) behaviour associated to the \lq effectiveness' (\lq non-effectiveness') of the catastrophe locus  is sketched in the upper (lower) part of in Fig. \ref{fig_hysteresis} for a cyclic variation in the control parameters.

The points $\mathsf{B}_1$ (magenta  in Fig. \ref{fig_hysteresis}, left) common to both $\mathcal{C}^{(+)}_P$ and $\mathcal{C}^{(-)}_P$  subsets are present only for effective sets and always correspond to supercritical pitchfork bifurcations as the limit case displaying a  non-snapping elastica \cite{cazzolli}. Contrarily, the sharp corner points $\mathsf{B}_2$ (yellow in Fig. \ref{fig_hysteresis}, left) within the subsets $\mathcal{C}^{(-)}_P$ or $\mathcal{C}^{(+)}_P$ are
	associated with subcritical pitchfork bifurcations displaying high-energy release for the snapping elastica.

Because of its generality, the present theoretical framework can be exploited to design a specific \emph{elastica catastrophe machine} by particularizing the kinematic relations $X_l(\bp,\bq), Y_l(\bp,\bq),$ and $\Theta_l(\bp,\bq)$. 
	Within the infinite set of possible \emph{elastica catastrophe machines}, as evidence of feasibility, two specific families are proposed and investigated in the next section, showing that catastrophe locus can be attained with peculiar properties by tuning the design parameters vector $\bq$. More specifically, the catastrophe locus $\mathcal{C}_P$  of the \emph{elastica catastrophe machine} might exhibit a number of bifurcation points not necessarily equal to four. Indeed, such multiplicity can vary here because coincident with the number, variable through the design parameters,  of intersections of the elastica machine set $\mathcal{E}_K$ (which is in general not a plane) with the snap-back set $\mathcal{S}_K$ with $\theta_S=0$ (points $\mathsf{B}_1$) or $\theta_A=0$ (points $\mathsf{B}_2$). Even more unusual, the catastrophe locus $\mathcal{C}_P$ may substantially vary its non-convexity, differently from the classical machines. This property is fundamental for crossing bifurcation points at high-energy release. Indeed, convexity facilitates reaching  bifurcation points, otherwise confined within acute angles in the classical machines. Because an analytical proof is awkward, with the purpose to evaluate the convex property of the catastrophe locus $\mathcal{C}_P$,  the convexity measure $\mathsf{C}(\mathcal{C}_P)$ is introduced \cite{convex}
	\beq
	\label{conv}
	\mathsf{C}(\mathcal{C}_P)=\dfrac{\text{Area}(\mathcal{C}_P)}{\text{Area}(\text{\textbf{CH}}(\mathcal{C}_P))}.
	\eeq
In eqn. (\ref{conv})  \textbf{CH}($\mathcal{C}_P$) is the \textit{convex hull} of $\mathcal{C}_P$, namely the smallest convex set including the shape of the catastrophe locus. The convexity measure $\mathsf{C}$ ranges between 0 and 1, being equal to 1 if and only if the planar shape is convex.

\section{ Two families of \textit{elastica catastrophe machines} }\label{families}

\begin{figure}[!htb]
\begin{center}
\includegraphics[width=160mm]{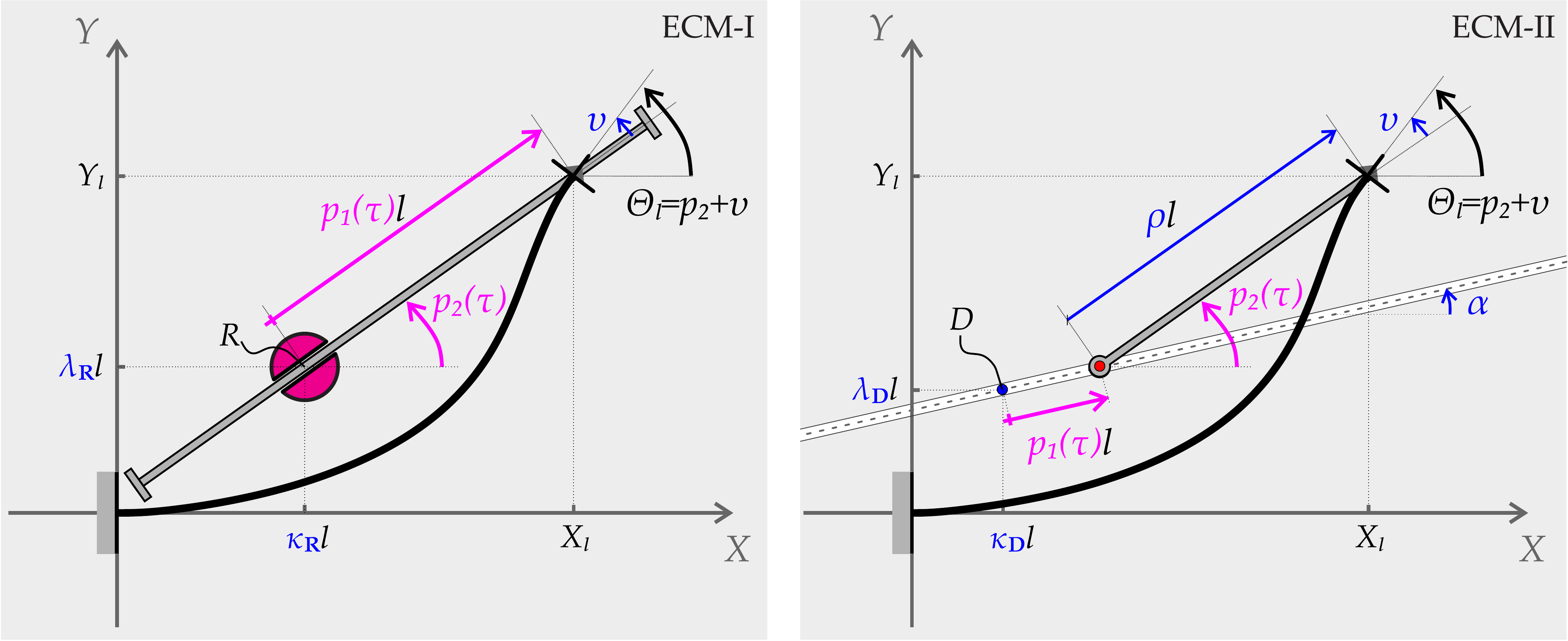}
\end{center}
\caption{\footnotesize{The two proposed families of \emph{elastica catastrophe machine}: ECM-I (left) and ECM-II (right). An inextensible elastic rod of length $l$ has an end with a fixed position and rotation. 
The kinematics of the elastic rod, considered within the plane $X-Y$,  is ruled by the configuration of an external rigid bar defined through two control parameters $p_1$ and $p_2$.  The design parameter vectors $\bq^ {I}=\{\kappa_R,\lambda_R,\upsilon\}$ and  $\bq^{II}=\{\kappa_D,\lambda_D, \alpha,\rho,\upsilon\}$ define respectively a family of \emph{elastica catastrophe machines} ECM-I and ECM-II.
}}
\label{machines0}
\end{figure}

With the purpose to explicitly evaluate catastrophe loci generated by \emph{elastica catastrophe machines}, the two  families ECM-I (Fig. \ref{machines0}, left) and ECM-II (Fig. \ref{machines0}, right) are considered and investigated by means of the general theoretical framework presented in the previous section. The elastica's end $s=l$ is considered attached to an external rigid bar, which configuration is defined by the two control parameters $p_1$ and $p_2$.
In both ECM-I and ECM-II the control parameter $p_2$ is taken coincident with the rigid bar rotation, so that, introducing the design angle parameter $\upsilon$ between the rigid bar and the elastica end tangent, the rotation $\Theta_l$ imposed at the final curvilinear coordinate is given by
\beq\label{thetaphiups}
\Theta_l(\bp,\bq)=p_2+\upsilon.
\eeq

For each one of the two proposed families, the dependence on the control parameters  is  specified for the physical coordinates $X_l(\bp,\bq)$ and $Y_l(\bp,\bq)$. Therefore, the respective inextensibility and machine sets, introduced in the previous Section with a general perspective, can be explicitly identified. Finally, the shape change of the corresponding catastrophe set and the achievement of \lq effective' catastrophe sets are disclosed with varying the design parameters vector $\bq$. 

\subsection{ The \textit{elastica catastrophe machine} ECM-I}
In the first family of catastrophe machine (Fig. \ref{machines0}, left), the external rigid bar is constrained by a sliding sleeve, centered at the fixed point $R=(\kappa_R l; \lambda_R l)$ and whose inclination with respect to the $X$-axis corresponds to the control parameter $p_2$. By sliding the  rigid bar, the distance $p_1 l$ between the elastica end $s=l$ and the sliding sleeve rotation center is ruled by the control parameter $p_1$, so that the coordinates of the elastica's end are
\beq
\label{kinemECM-I}
\overline{X}_l(\bp,\bq^I)=\left(\kappa_R+p_1\cos p_2\right) l,\qquad
\overline{Y}_l(\bp,\bq^I)=\left(\lambda_R+p_1 \sin p_2\right) l,
\eeq
with  $p_1$ restricted to positive values ($p_1>0$)\footnote{
	It is noted that the catastrophe sets related to negative values (disregarded here) of $p_1$  can be obtained from those restricted to positive values, being the configuration corresponding to a control vector $\bp=\{p_1^\flat, p_2^\flat\}$ and a design vector $\bq^I=\{\kappa_R^\flat, \lambda_R^\flat, \upsilon^\flat\}$  the same of that corresponding to $\bp=\{-p_1^\flat, p_2^\flat+\pi\}$ and $\bq^I=\{\kappa_R^\flat, \lambda_R^\flat, \upsilon^\flat-\pi\}$.
}
and the control parameters vector has length $N=3$ and is given by
\beq
\bq^I=\{\kappa_R, \lambda_R, \upsilon\}.
\eeq

The different relations connecting the configuration representation through control parameters, primary kinematical quantities and physical coordinates can be  derived  from the explicit kinematical rules (\ref{thetaphiups}) and (\ref{kinemECM-I}). These are reported in the Supplementary Material (Section \ref{appECM-I}).

The \lq elastica machine set' $\mathcal{E}_C$ is defined in the control parameters plane $p_1-p_2$ as the intersection of the inextensibility set $\mathcal{I}_C$, provided by the inextensibility condition (\ref{distanced})  as
\beq\label{inexECMI}
\mathcal{I}_C=\left\{\bp:\,\,
p_1^2+ 2p_1 \left(\kappa_R \,\cos p_2+ \lambda_R  \sin p_2\right)
+{\kappa_R }^2+{\lambda_R}^2 \leq 1\right\},
\eeq
with the set machine domain $\mathcal{M}_C$, eqn (\ref{limitazionesymasym}), expressed  as
\beq\label{macECMI}
\mathcal{M}_C=\left\{\bp:\,\,
\left|\frac{p_2+ \upsilon}{2}\right|<\pi,
\left|\frac{p_2+ \upsilon}{2}-\arctan\left(\frac{\lambda_R+p_1\sin p_2}{\kappa_R+p_1 \cos p_2}\right)\right|<\pi\right\}.
\eeq

The different sets in the primary kinematical space and in the physical plane can be obtained from the respective projections of $\mathcal{I}_C$ (\ref{inexECMI}) and $\mathcal{M}_C$ (\ref{macECMI}) by means of equations (\ref{maicanta})-(\ref{maicantafin}). 

\begin{figure}[!h]
	\begin{center}
		\includegraphics[width=180mm]{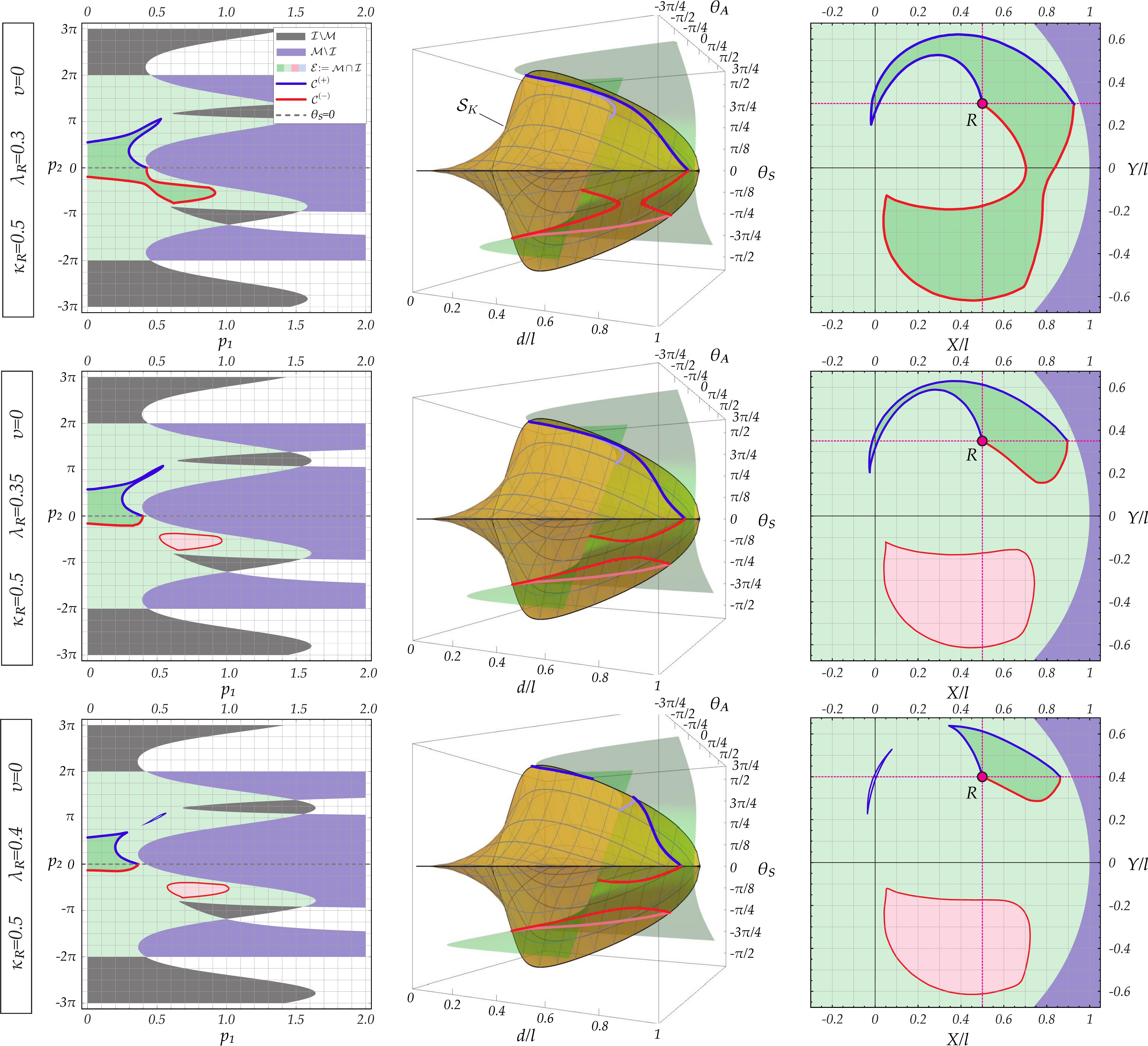}
	\end{center}
	\caption{
		Inextensibility $\mathcal{I}$, machine $\mathcal{M}$, elastica $\mathcal{E}$, and  catastrophe $\mathcal{C}$ sets  within the three different spaces (from the left to right: control plane, primary kinematical space, physical plane) for three ECMs-I with $\kappa_R=0.5$, $\upsilon=0$ and $\lambda_R=\{0.3,0.35,0.4\}$ increasing from the first to the third line. The yellow surface appearing in the primary kinematical space $\{d/l,\theta_A,\theta_S\}$ is the snap-back set $\mathcal{S}_K$. The portions $\mathcal{C}^{(+)}$/$\mathcal{C}^{(-)}$ of the \lq catastrophe sets' are reported as  blue/red lines. The effective \lq catastrophe sets' are reported as thick lines and are those defined by the union of $\mathcal{C}^{(+)}$ with $\mathcal{C}^{(-)}$ (where the superscript is related to the sign of the ends' curvature of the snapping configuration). The shifting of the rotation centre $R$ parallel to the $Y$-axis provides a change of the catastrophe locus, with the realization of   possible non-effective sets (closed thin lines drawn with only one colour).
	}
	\label{fig_detachECM-I}
\end{figure}

\begin{figure}[!htb]
	\begin{center}
		\includegraphics[width=180mm]{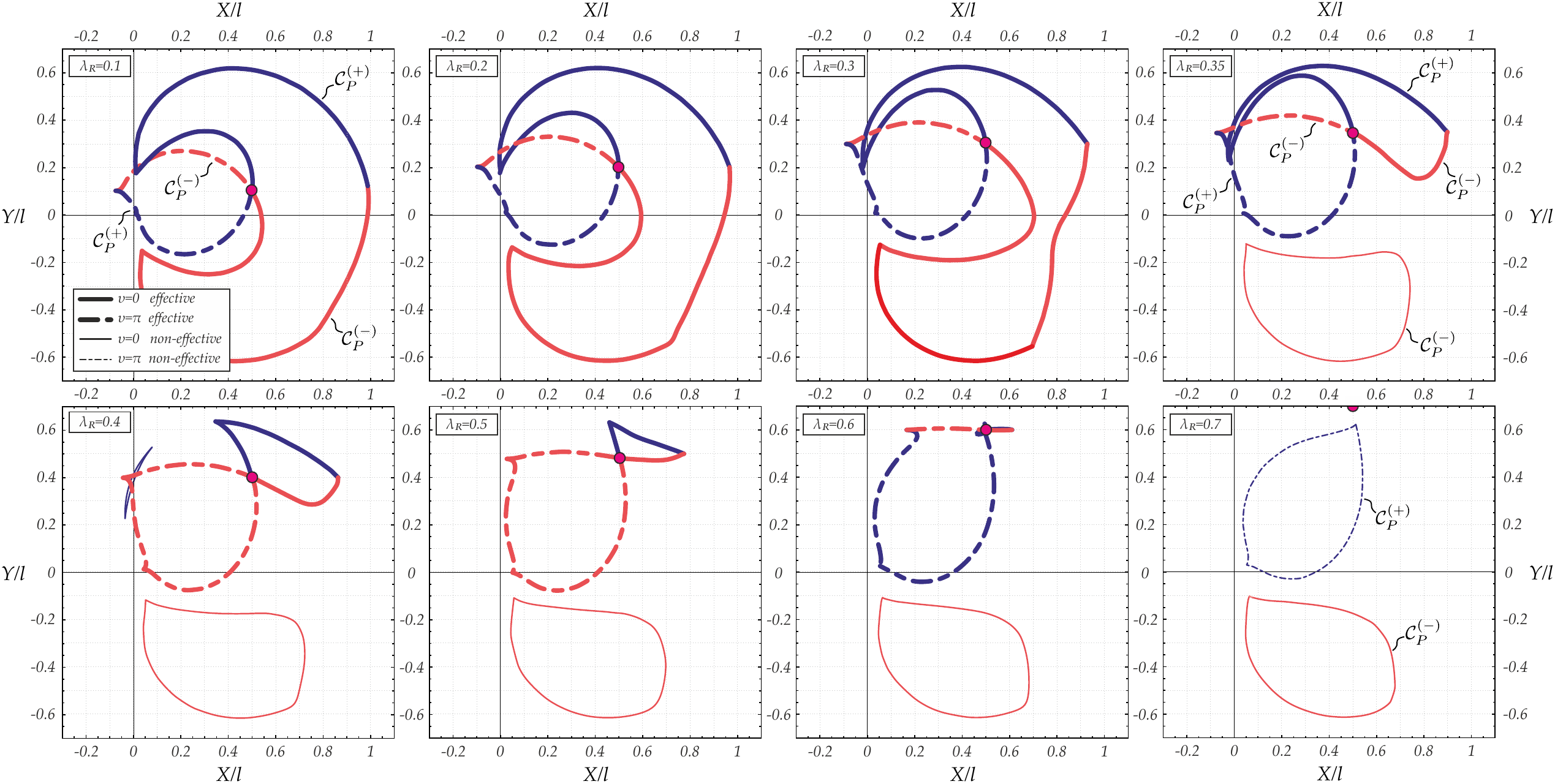}
	\end{center}
	\caption{\footnotesize{Catastrophe loci in the physical plane $X-Y$ for ECM-I with $\kappa_R=0.5$, $\lambda_R=\{0.1,0.2,0.3,0.35,0.4,0.5,0.6,0.7\}$}, and $\upsilon=\{0,1\}\pi$. The red spots identify the position rotation center $R$ of the rigid bar. Effectiveness and non-effectiveness of the catastrophe loci is distinguished through thick and thin coloured lines, respectively.}
	\label{ECMI_kR05}
\end{figure}

\begin{figure}[!htb]
	\begin{center}
		\includegraphics[width=180mm]{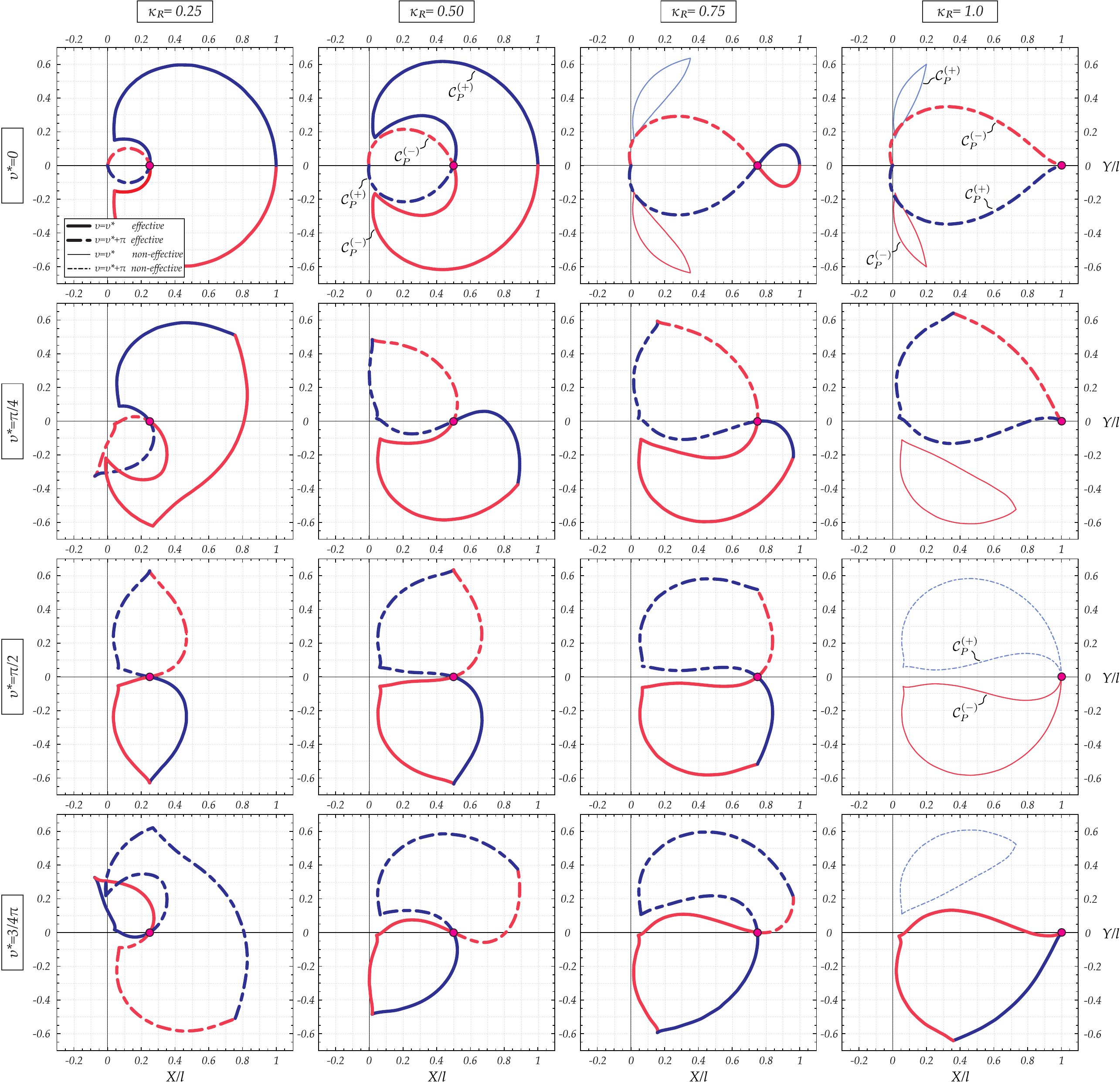}
	\end{center}
	\caption{\footnotesize{As for Fig. \ref{ECMI_kR05}, but for $\lambda_R=0$, $\kappa_R=\{0.25,0.5,0.75,1\}$ and $\upsilon=\{0,1/4,1/2,3/4,1,5/4,3/2,7/4\}\pi$}. The parameter $\upsilon^*$ is introduced to represent the two cases $\upsilon=\upsilon^*$ and $\upsilon=\upsilon^*+\pi$ in the same plot. The position rotation center R of the rigid bar is identified as red spots.}
	\label{ECMI_LR00_2}
\end{figure}

The  $\mathcal{I}$, $\mathcal{M}$, $\mathcal{E}$, and $\mathcal{C}$  sets are reported in 
Fig. \ref{fig_detachECM-I}  for the control parameters $\kappa_R=0.5$, $\upsilon=0$ and $\lambda_R=\{0.3,0.35,0.4\}$. Therefore, the considered ECMs-I  differ only in the position of the rigid bar rotation center $R$, slightly moving up from the first to the third line. In the Figure, three different spaces are depicted: the control plane (left column), the primary kinematical space (central column, also containing the snap-back surface $\mathcal{S}_K$), and  the physical plane (right column). 
The catastrophe set $\mathcal{C}_K$ is evaluated within the primary kinematical space as the curve defined by the intersection of two surfaces, representing  the snap set $\mathcal{S}_K$   and the elastica set $\mathcal{E}_K$. The obtained catastrophe curve has projections $\mathcal{C}_C$ and $\mathcal{C}_P$ within the control and physical planes as planar curves. The positive and negative sign of ends' curvature related to the configuration at snapping is highlighted along the catastrophe curves $\mathcal{C}$ with blue ($\mathcal{C}^{(+)}$) and red  ($\mathcal{C}^{(-)}$) colour, respectively. How the catastrophe set changes with increasing the design parameter $\lambda_R$ may be appreciated from the figure. In particular, the sets of coordinates corresponding to two stable equilibrium configurations are given by the union of one ($\lambda_R=0.3$), two ($\lambda_R=0.35$), and three ($\lambda_R=0.4$) simply connected domains in the physical plane. However, for each of the three cases, only one of these simply connected domains (drawn as a thick line) provides hysteresis when crossed  during a cyclic path.
When existing ($\lambda_R=\{0.35, 0.4\}$), the other simply connected domains have boundary (drawn with thin line) for which snapping occurs only for a positive \emph{or} for a negative sign of the ends' curvature, so that \emph{no more than one} snap (and therefore \emph{no hysteresis}) can be related to these during a cyclic path.  The existence and properties of these simply connected domains are strictly dependent on the selected design vector. With this regard, considering\footnote{Symmetry and periodicity properties  of expressions (\ref{thetaphiups}) and (\ref{kinemECM-I}) in the design parameters $\lambda_R$ and $\upsilon$ define symmetry properties for the catastrophe sets. More specifically, a change in the sign of both the design parameters $\lambda_R$ and $\upsilon$ provides a mirroring of the catastrophe set (with respect to the $X$ axis within the physical plane and with respect to the $p_2$ within the control plane) and a change in  sign for the related curvature displaying  snapping. An increase of $2k \pi$ ($k\in\mathbb{Z}$) in $\upsilon$
	provides a shifting of $-2k \pi$ of the catastrophe set with respect to the $p_2$ axis  within the control plane and no change within the physical plane.
} 
 only non-negative values of $\lambda_R$ and  $\upsilon\in[0,2\pi)$ and restricting the attention to the physical plane representation,  the influence of the design parameters is shown through the following  effective and non-effective catastrophe sets:
\begin{itemize}
	\item for $\kappa_R=0.5$, $\upsilon=\{0,1\}\pi$, and $\lambda_R=\{0.1,0.2,0.3,0.35,0.4,0.5,0.6,0.7\}$ in Fig. \ref{ECMI_kR05}, showing that the \emph{elastica catastrophe machine} is non-effective for both the considered values of $\upsilon$ when $\lambda_R=0.7$, while effective in the remaining cases;
	\item for $\kappa_R=\{0.25,0.5,0.75,1\}$, $\upsilon=\{0,1/4,1/2,3/4,1,5/4,3/2,7/4\}\pi$, and $\lambda_R=0$ in Fig. \ref{ECMI_LR00_2}, showing that the \emph{elastica catastrophe machine} is non-effective for $\kappa_R=1$ when $\upsilon=\{0,1/4,1/2,3/2,7/4\}\pi$, while effective in the remaining cases.
\end{itemize}  

Dramatic changes in the projection of the catastrophe set within the physical plane can be observed from these figures, as the result of changing the design parameter vectors $\bq$. In particular, a loss of symmetry in the catastrophe locus occurs when $\upsilon\neq\{0,\pi\}$ or $\lambda_R\neq0$. Furthermore, the catastrophe loci that may be generated by ECM-I encompass a large variety of shapes, very different from those related to the classical catastrophe machines. In particular, the following new features of the catastrophe sets are found:
\begin{itemize}\item \textit{Variable number of bifurcation points.}
The effective catastrophe loci, reported for different values of $\bq$  in Figs. \ref{ECMI_kR05} and \ref{ECMI_LR00_2}, display a  variable number of bifurcation points depending on the number of intersections of   $\mathcal{E}_K$ with $\mathcal{S}_K$ for $\theta_S=0$ and $\theta_A=0$. In Fig. \ref{ECMI_kR05}, the reported effective sets have two (e.g. second row, first column,  $\upsilon=0$) or four bifurcation points (e.g. first row, second column,  $\upsilon=0$). In Fig. \ref{ECMI_LR00_2}, the number of bifurcation points can be equal to one (e.g. first row, first column, $\upsilon=\pi$), two (e.g. second row,  second column, $\upsilon=\pi/4$), three (first row,  first column,  $\upsilon=0$) or four (second row, first column, $\upsilon=\pi/4$);

\item \textit{Convex measure of the catastrophe locus $\mathcal{C}_P$.}
In Fig. \ref{ECMI_LR00_2}, the catastrophe sets for $\{\upsilon=\pi, \, \kappa_R=0.25\}$ (first row, first column) and for $\{\upsilon=0, \, \kappa_R=0.75\}$  (first row, third column) have convex measure approaching the unit value, $\mathsf{C}\simeq 1$. The catastrophe set for $\{\upsilon=\pi, \, \kappa_R=0.5\}$ (first row, second column) has $\mathsf{C}=0.9997$ while that for $\{\upsilon=\pi, \, \kappa_R=0.75\}$  (first row, third column) has $\mathsf{C}=0.998$. Finally,  the catastrophe sets for $\upsilon=0$ and $\lambda_R=\{0.3,0.35,0.4\}$ shown in Figs. \ref{fig_detachECM-I} and  \ref{ECMI_kR05} have $\mathsf{C}=\{0.5707,\,0.4475,\,0.9402\}$.
\end{itemize}

The recommendations about how to select the initial values of the control parameters $\bp(\tau_0)$ to belong to the inextensibility set $\mathcal{I}_C$  are included in Sect. \ref{appECM-I} of the Supplementary Material. Some considerations drawn for ECM-I under the two special conditions for the position center $R$ are respectively reported in Sects. \ref{appB_11} and \ref{appB_12}) of the Supplementary Material.
\\Similarly to the classical Zeeman machine, all the sharp corner points, when not coincident with the rotation centre $R$, correspond to pitchfork bifurcations.

\subsection{The \textit{elastica catastrophe machine}  ECM-II}

In the catastrophe machine ECM-II (Fig. \ref{machines0}, right) the rigid bar of fixed length $\rho l$ can rotate and has one end constrained to slide along a straight line, inclined at an angle $\alpha$ with respect to the $X$-axis.  The center of rotation of the rigid bar is at a  controlled distance $p_1 l$ from a  fixed point $D$, of coordinates $\{X_D,Y_D\}=\{\kappa_D,\lambda_D\}l$, located on the straight line. By controlling the inclination $p_2$ and the distance $p_1 l$ of the movable rotation center of coordinates $\{\kappa_D+p_1\cos \alpha, \lambda_D+p_1 \sin \alpha\}l$ from the reference point $D$, the elastica end $s=l$ has coordinates
\beq
\label{kinemECM-II}
\overline{X}_l(\textbf{p},\textbf{q}^{II})
=\left(\kappa_D+p_1\cos\alpha+\rho \cos p_2\right) l,\qquad
\overline{Y}_l(\textbf{p},\textbf{q}^{II})=\left(\lambda_D+p_1 \sin\alpha+\rho \sin p_2\right) l,
\eeq
while the design parameters  vector (of length $N=5$) for ECM-II is
\beq\label{macchII}
\bq^{II}=\{\kappa_D, \lambda_D, \alpha,\rho, \upsilon\}.
\eeq

The equations relating the control parameters, the primary kinematical quantities and the physical coordinates are obtained through the equations (\ref{thetaphiups}) and (\ref{kinemECM-II}) and reported in the Supplementary Material (Section \ref{appECM-II}).

The control parameters vector $\bp$ for ECM-II is restricted to the set $\mathcal{E}_C$, given by the intersection of the inextensibility set $\mathcal{I}_C$,
\beq\label{inexECMII}
\resizebox{1\textwidth}{!}{$
	\mathcal{I}_C=\left\{\bp
	:p_1^2+{\rho}^2+ 
	2\left[
	p_1 \rho \cos({\alpha}- p_2)
	+\kappa_D\left(p_1 \cos{\alpha}+\rho \cos{p_2}\right) 
	+\lambda_D\left(p_1\sin{\alpha}+\rho \sin{p_2}\right) 
	\right]
	+{\kappa_D}^2
	+{\lambda_D}^2 \leq 1\right\},
	$}
\eeq
with the machine domain $\mathcal{M}_C$
\beq
\mathcal{M}_C=\left\{\bp
:\left|\frac{p_2+ \upsilon}{2}\right|<\pi,
\left|\frac{p_2+ \upsilon}{2}-\arctan\left(\frac{\lambda_D+p_1\sin\alpha+\rho \sin p_2}{\kappa_D+p_1\cos\alpha+\rho \cos p_2}\right)\right|<\pi\right\}.
\eeq

In order to have a non-null elastica machine set  $\mathcal{E}$,  the first four design parameters are constrained to satisfy the following inequality
\beq
\left|
\kappa_D \sin\alpha- \lambda_D \cos\alpha
\right|
-\rho<1.
\eeq

From eqn (\ref{kinemECM-II}), it may also be noted that two different control parameters vectors $\bp^{\flat}$ and $\bp^{\sharp}$ provide the same end's coordinates $\{X_l(\bp^{\flat},\bq))=X_l(\bp^{\sharp},\bq),
Y_l(\bp^{\flat},\bq))=Y_l(\bp^{\sharp},\bq)\}$ when
\beq
p_1^{\sharp}=p_1^{\flat}+ 2 \rho \cos(p_2^{\flat}-\alpha),
\qquad
p_2^{\sharp}=\pi+2\alpha-p_2^{\flat}.
\eeq
Therefore, in addition to the natural  multiplicity due to the angular periodicity in the physical angle $\Theta_l$,  the same position $X_l,Y_l$ in the physical plane is associated to two physical angles $\Theta_l(\bp^{\sharp},\bq)\neq\Theta_l(\bp^{\flat},\bq)$ not differing by  $2k\pi$ ($k\in \mathbb{Z}$), namely
\beq
\Theta_l(\bp^{\sharp},\bq)=\pi+2(\alpha+\upsilon)-\Theta_l(\bp^{\flat},\bq).
\eeq
Due to this additional multiplicity, it is instrumental to analyze the behaviour of ECM-II through the analysis of two machine subtypes, ECM-IIa and ECM-IIb, having  the control parameter $p_2$ (ruling the rigid bar rotation)  restricted to specific sets $p_2^{(a)}$ and $p_2^{(b)}$, 
\begin{equation}
\label{vara}
\begin{array}{lll}
\mbox{ECM-IIa}:\,\,\,
p_2^{(a)}\in\ds\bigcup_{k\in\mathbb{Z}}\left[-\dfrac{\pi}{2}+\alpha+2k\pi,\dfrac{\pi}{2}+\alpha+2k\pi\right],\\[4mm]
\mbox{ECM-IIb}:\,\,\,
p_2^{(b)}\in\ds\bigcup_{k\in\mathbb{Z}}\left[\dfrac{\pi}{2}+\alpha+2k\pi,\dfrac{3\pi}{2}+\alpha+2k\pi\right].
\earr
\end{equation}

\begin{figure}[!htb]
	\begin{center}
		\includegraphics[width=0.9\textwidth]{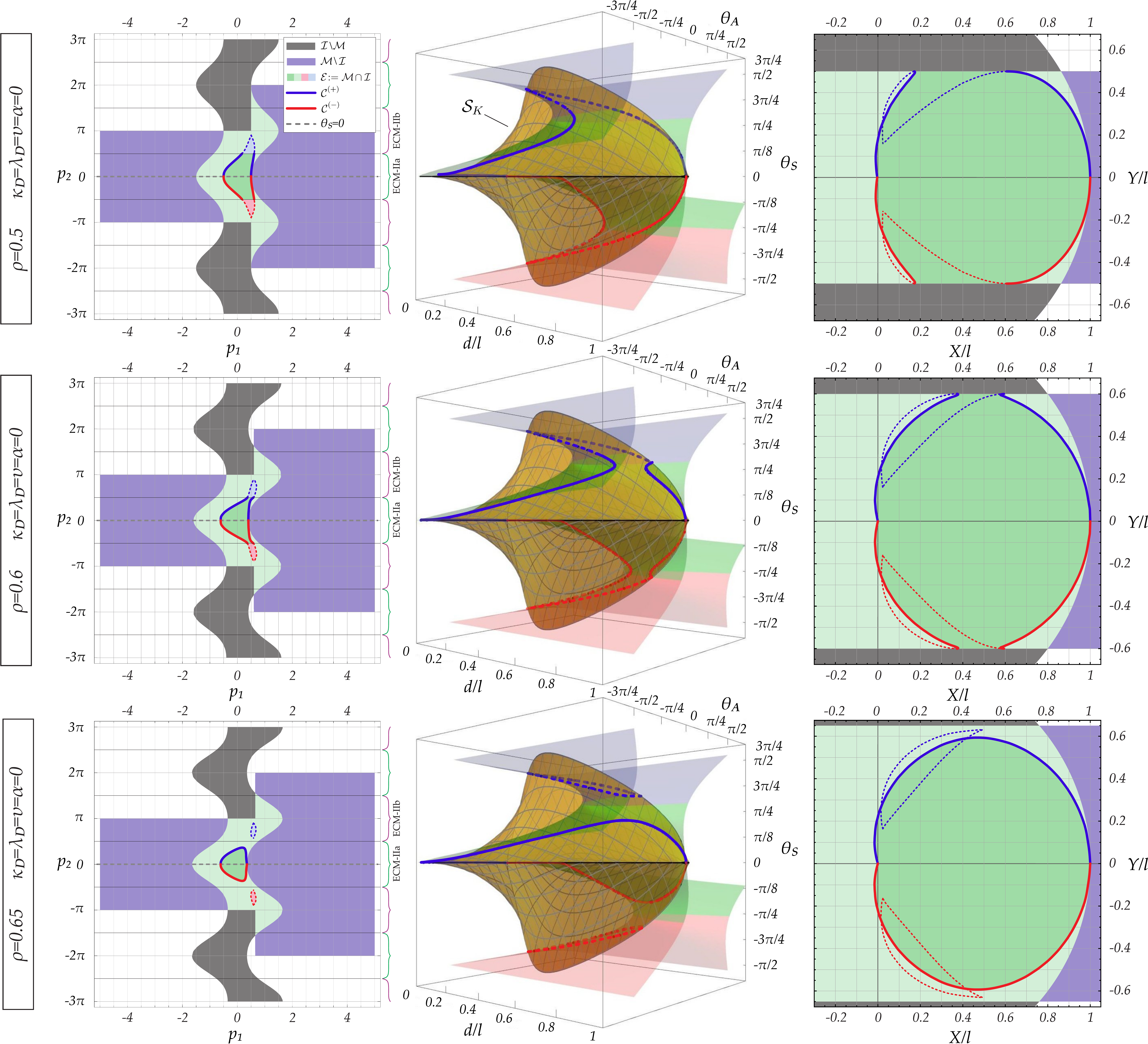}
	\end{center}
	\caption{
		As for Fig. \ref{fig_detachECM-I}, but for three ECMs-II with $\kappa_D=\lambda_D=\upsilon=\alpha=0$  and $\rho=\{0.5,0.6,0.65\}$ increasing from the first to the third line.  The portions $\mathcal{C}^{(+)}$/$\mathcal{C}^{(-)}$ of the \lq catastrophe sets' are reported as  blue/red lines, continuous for ECM-IIa and dashed for ECM-IIb. For the reported cases, the effective \lq catastrophe sets' are related only to ECM-IIa.
		The increase in the rigid bar parameter length $\rho$ provides a change of the catastrophe locus, as the detachment from the elastica machine set boundary in the case $\rho=0.65$.
	}
	\label{fig_detachECM-II}
\end{figure}

The inextensibility $\mathcal{I}$, the machine $\mathcal{M}$, the elastica $\mathcal{E}$ and catastrophe $\mathcal{C}$ sets for ECM-II are shown in Fig. \ref{fig_detachECM-II}  within the control plane (left column), primary kinematical space (central column), and the physical plane (right column) for $\kappa_D=\lambda_D=\alpha=\upsilon=0$ and $\rho=\{0.5,0.6,0.65\}$, with increasing values from the upper to the lower line.
The portions $\mathcal{C}_P^{(+)}$/$\mathcal{C}_P^{(-)}$ of catastrophe loci  are reported as blue/red lines, continuous for ECM-IIa  and dashed for ECM-IIb. In the figure, the blue/red line defines configurations for which elastica with positive/negative ends curvature snaps.
The thick line identifies an effective catastrophe locus while a thin line a non-effective one, so that, in Fig. \ref{fig_detachECM-II}, the catastrophe  sets of EMC-IIa are all effective while those of ECM-IIb are not. 
It is evident that for $\rho=\{0.5,0.6\}$ the catastrophe sets of ECM-IIa and ECM-IIb have in common their end points (also at the boundary of $\mathcal{I}$/$\mathcal{M}$ in the physical plane) while for $\rho=0.65$ the two machine subtypes do not share any point with each other.

\begin{figure}[!htb]
	\begin{center}
		\includegraphics[width=0.75\textwidth]{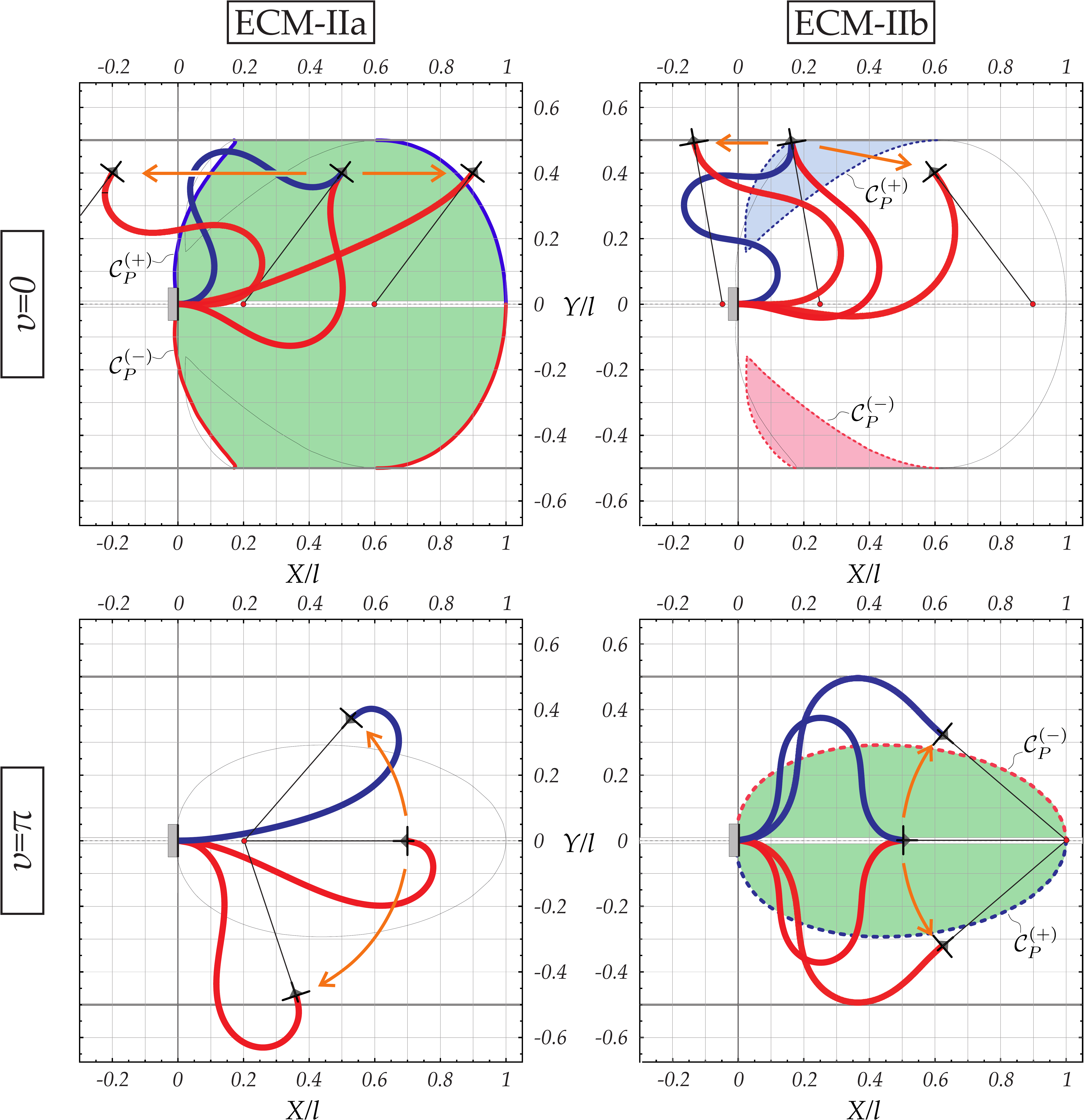}
	\end{center}
	\caption{Catastrophe sets of ECM-IIa and ECM-IIb with $\kappa_D=\lambda_D=\alpha=0$, $\rho=0.5$, and $\upsilon=0$ (upper part) or $\upsilon=\pi$ (lower part). While ECM-IIa is effective for $\upsilon=0$, ECM-IIb is effective for $\upsilon=\pi$. Deformed configurations are also displayed for some specific end's position, highlighting equilibrium multiplicity (when existing).}
	\label{fig_detachECM-II_2}
\end{figure}

The catastrophe sets of ECM-IIa and ECM-IIb for $\kappa_D=\lambda_D=\alpha=\upsilon=0$ and $\rho=0.5$ shown in Fig. \ref{fig_detachECM-II} (upper part, right) are also  reported separately for the two machine subtypes in Fig. \ref{fig_detachECM-II_2} (upper part). This latter representation contains also some elasticae,  highlighting the two possible equilibrium configurations for a state within the bistable region and the only possible configuration after crossing the catastrophe locus. The catastrophe sets are also reported in the lower  part of  Fig. \ref{fig_detachECM-II_2} for a machine with the same design parameters of that considered in the upper part of  Fig. \ref{fig_detachECM-II_2} except for the value of $\upsilon$, taken as $\upsilon=\pi$ (instead of $\upsilon=0$).

\begin{figure}[!htb]
	\begin{center}
		\includegraphics[width=1\textwidth]{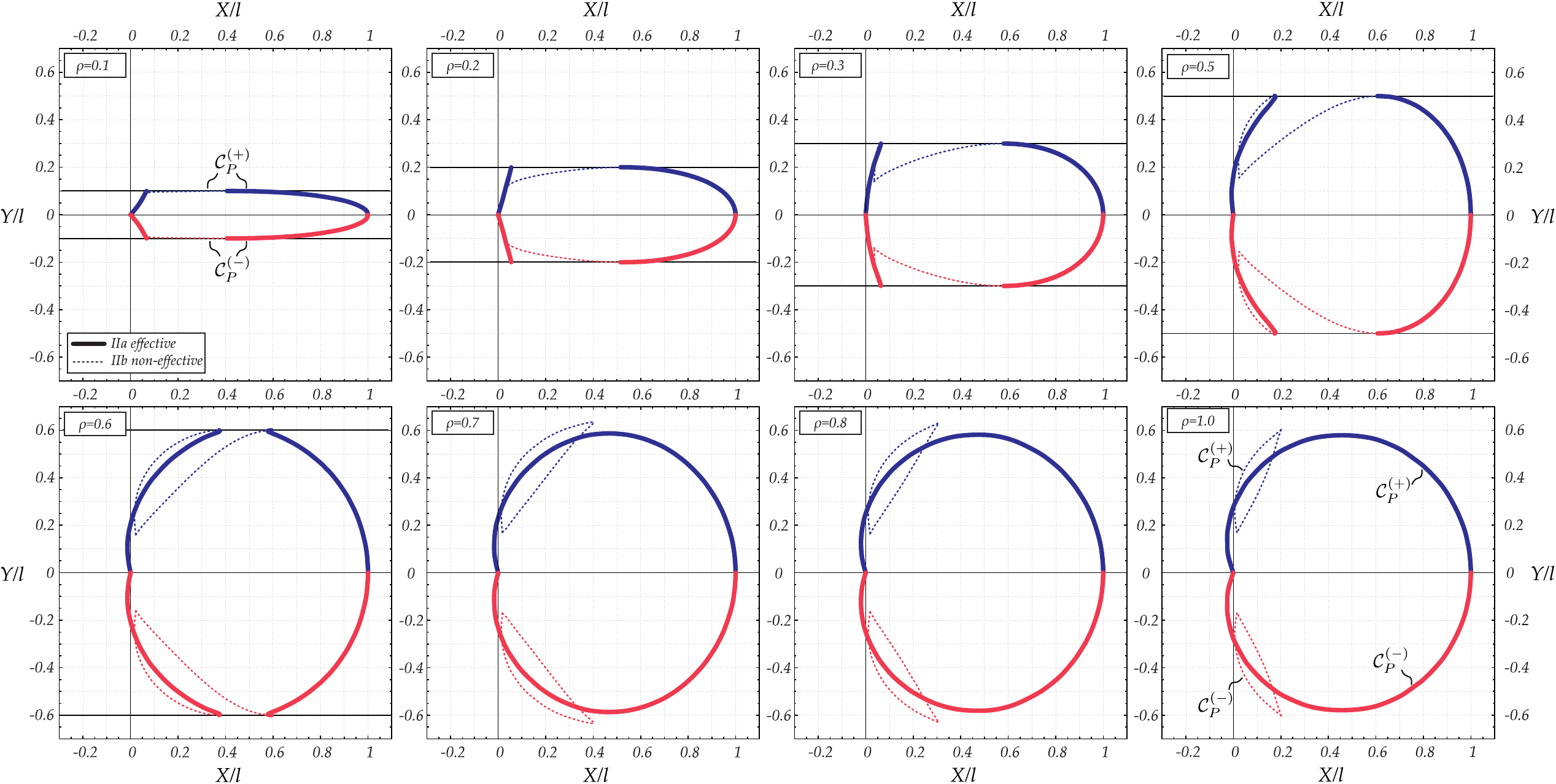}
	\end{center}
	\caption{\footnotesize{Catastrophe loci in the physical plane $X-Y$ for ECM-II with $\kappa_D=\lambda_D=\alpha=\upsilon=0$ and $\lambda_D=\{0.1,0.2,0.3,0.5,0.6,0.7,0.8,1\}$. Sets related to ECM-IIa are represented by continuous lines while those releated to EMC-IIb by dashed lines. Effectiveness and non-effectiveness of the catastrophe loci is distinguished through thick and thin coloured lines, respectively.} }
	\label{ECMII_alpha0_nu0}
\end{figure}

\begin{figure}[!htb]
	\begin{center}
		\includegraphics[width=1\textwidth]{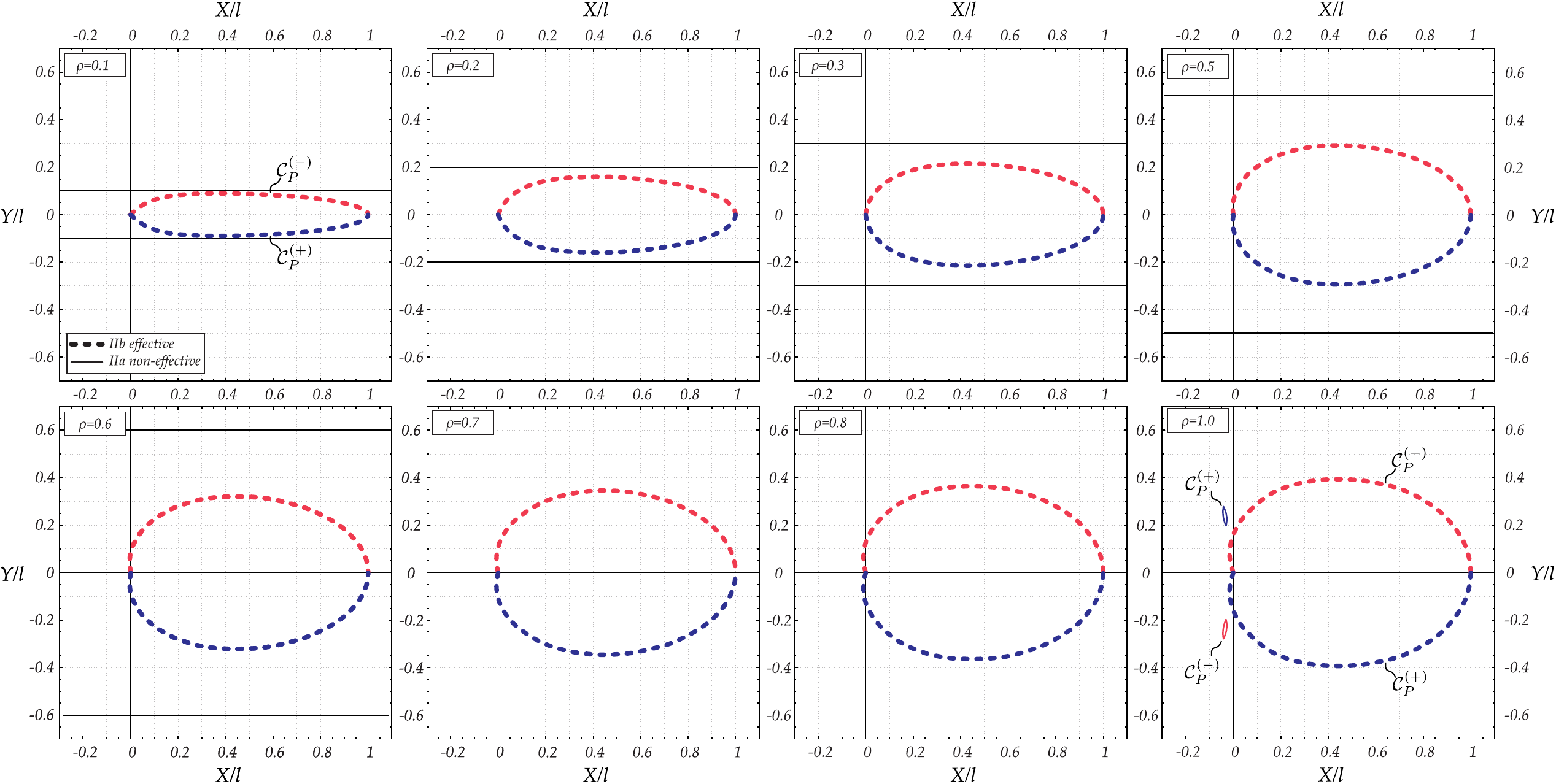}
	\end{center}
	\caption{\footnotesize{As for Fig. \ref{ECMII_alpha0_nu0}, but for $\upsilon=\pi$. }}
	\label{ECMII_alpha0_nupi}
\end{figure}

\begin{figure}[!htb]
	\begin{center}
		\includegraphics[width=0.9\textwidth]{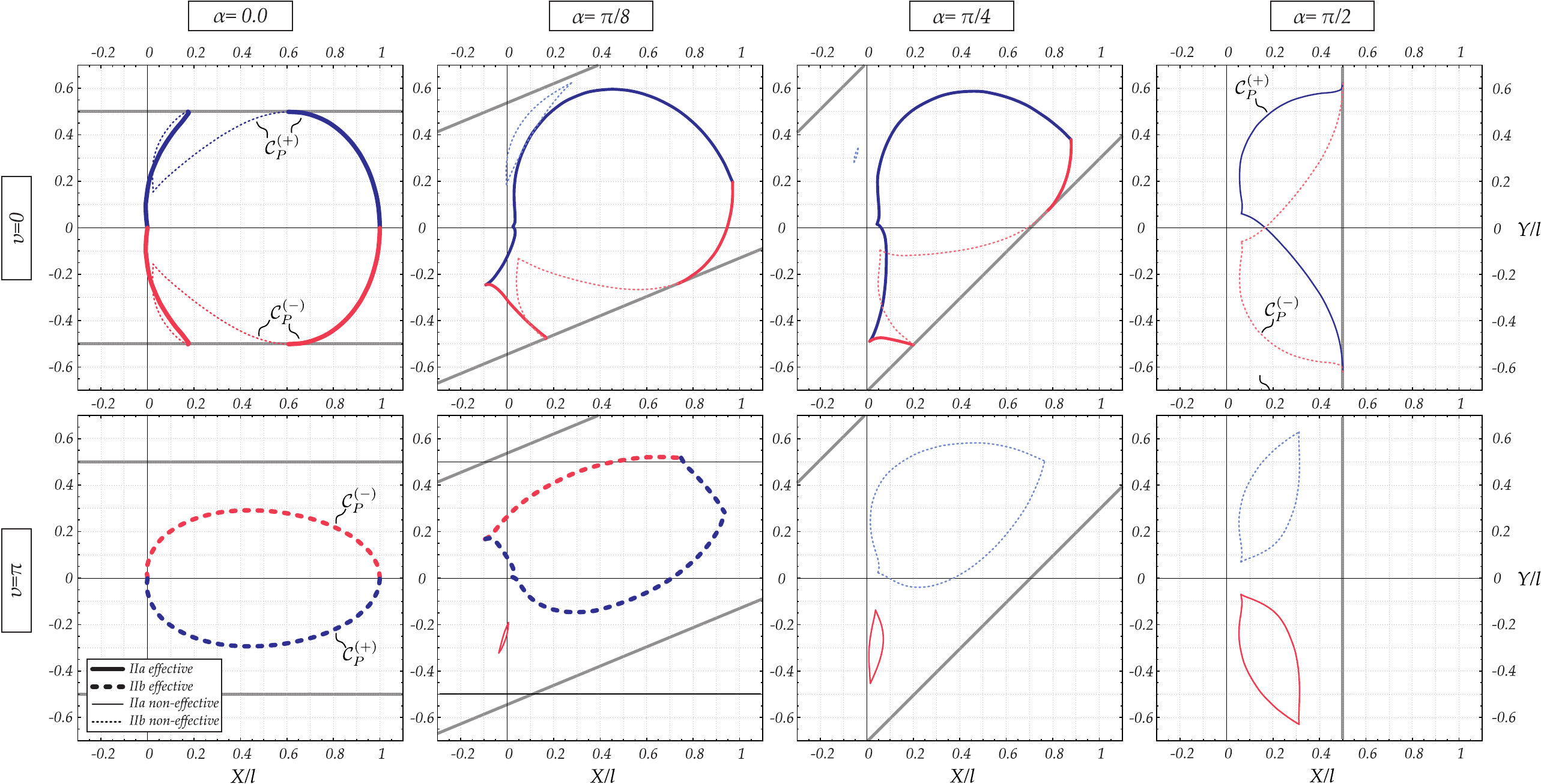}
	\end{center}
	\caption{\footnotesize{As for Fig. \ref{ECMII_alpha0_nu0}, but for $\kappa_D=\lambda_D=0$, $\rho=0.5$. A constant value of $\upsilon$ is considered for each line (first line $\upsilon=0$, second line $\upsilon=\pi$) and of $\alpha$ for each column ($\alpha=\{0,1/8,1/4,1/2\}\pi$, increasing from left to right).}}
	\label{ECMII_R05_alpha}
\end{figure}

\begin{figure}[!htb]
	\begin{center}
		\includegraphics[width=1\textwidth]{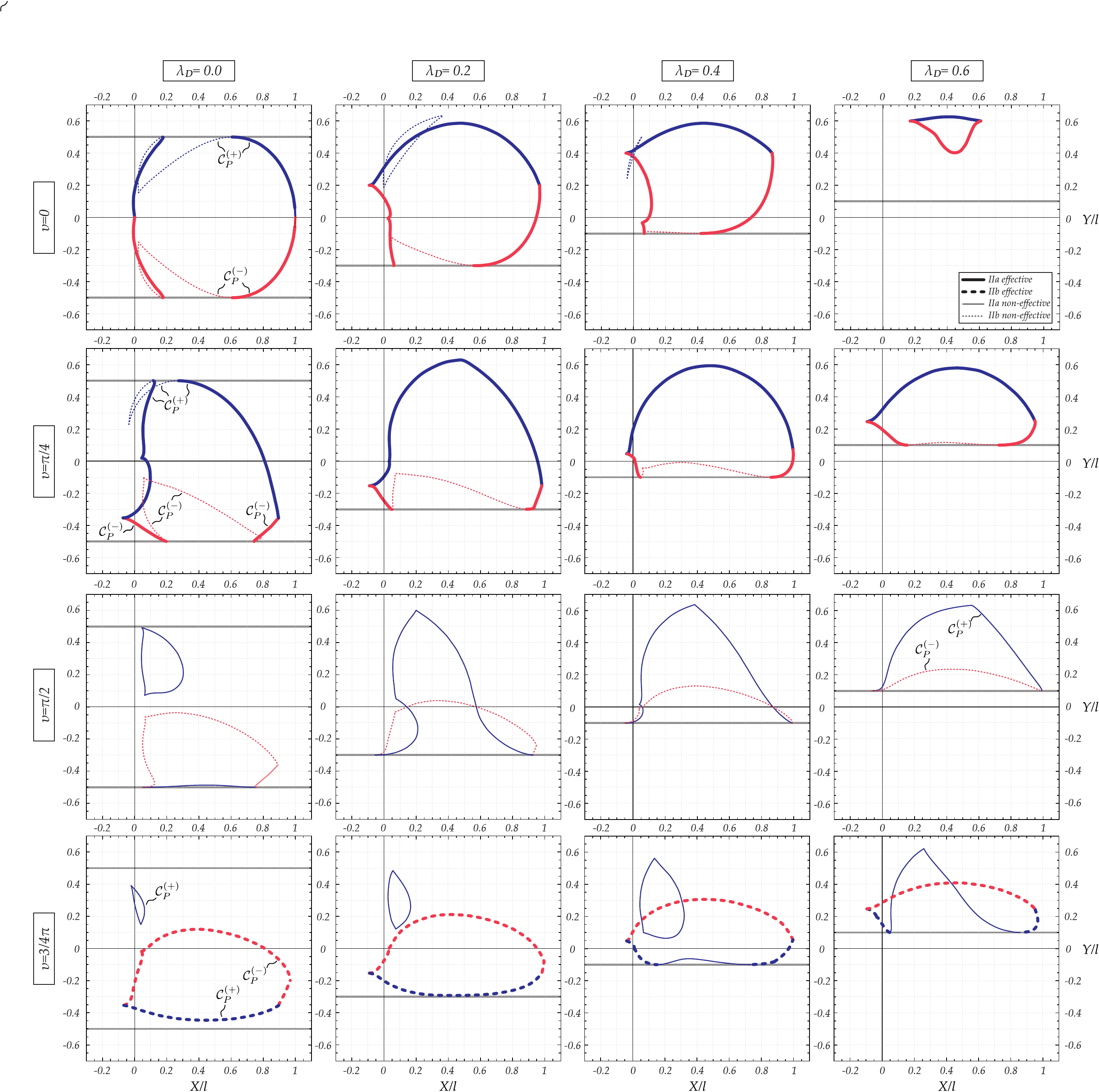}
	\end{center}
	\caption{\footnotesize{As for Fig. \ref{ECMII_alpha0_nu0}, but for $\kappa_D=\alpha=0$, $\rho=0.5$. A constant value of $\upsilon$ is considered for each line ($\upsilon=\{0,1/4,1/2,3/4\}\pi$, increasing from above to bottom) and of $\lambda_D$ for each column ($\lambda_D=\{0,0.2,0.4,0.6\}$, increasing from left to right).}}
	\label{ECMII_LDnu1}
\end{figure}

It can be appreciated that for the same design parameters no more than one of the two machine subtypes  is effective, namely ECM-IIa with $\upsilon=0$ and ECM-IIb with $\upsilon=\pi$. The remaining two machine subtypes are non-effective, but each of them  displays a different behaviour. 
ECM-IIb with $\upsilon=0$ has a non-effective catastrophe set so that only one snap may  occur during a continuous evolution while  ECM-IIa with $\upsilon=\pi$ has no catastrophe locus so that no snap is possible with this machine subtype.
Performing a parametric analysis of ECM-II (by varying the design parameters) it can be  concluded that  a principle of exclusion  about the effectiveness for the two subtypes of machine exists, namely, if ECM-IIa (or ECM-IIb) is an effective machine then ECM-IIb (or ECM-IIa) is not. This principle finds also evidence in the Figs. \ref{ECMII_alpha0_nu0}-\ref{ECMII_LDnu1}, referred to different design parameters vectors (restricted to $\upsilon\in[0,2\pi)$ due to periodicity) as follows:
\begin{itemize}
	\item for $\kappa_D=\lambda_D=\alpha=0$, $\rho=\{0.1,0.2,0.3,0.5,0.6,0.7,0.8,1\}$ and  $\upsilon=0$ in Fig. \ref{ECMII_alpha0_nu0} and for  the same values of  $\lambda_D$, $\kappa_D$, $\alpha$, and $\rho$ but $\upsilon=\pi$ in Fig. \ref{ECMII_alpha0_nupi},  showing that only ECM-IIa is effective in the former figure while only ECM-IIb is effective in the latter. In particular, ECM-IIa does not display any catastrophe set in all the cases in Fig.  \ref{ECMII_alpha0_nupi} (similarly to Fig. \ref{fig_detachECM-II_2}, bottom left) except when $\rho=1$;
	\item for $\kappa_D=\lambda_D=0$, $\rho=0.5$, $\upsilon=\{0,\pi\}$ and $\alpha=\{0,1/8,1/4,1/2\}\pi$ in Fig. \ref{ECMII_R05_alpha}, showing that only ECM-IIa is effective for $\upsilon=0$ and $\alpha=\{0,1/8,1/4\}\pi$,  only ECM-IIb is effective for $\upsilon=\pi$ and $\alpha=\{0,1/8\}\pi$, while no machine subtype is effective in the remaining cases;
	\item for $\kappa_D=\alpha=0$, $\rho=0.5$, $\lambda_D=\{0,0.2,0.4,0.6\}$ and $\upsilon=\{0,1/4,1/2,3/4\}\pi$  in Fig. \ref{ECMII_LDnu1}, showing that only ECM-IIa is effective for $\upsilon=\{0,1/4\}\pi$ for all the reported values of $\lambda_D$ and only ECM-IIb is effective for $\upsilon=3/4\pi$ for all the reported values of $\lambda_D$, while no machine subtype is effective in the remaining cases.
\end{itemize}

\begin{figure}[!htb]
	\begin{center}
		\includegraphics[width=0.7\textwidth]{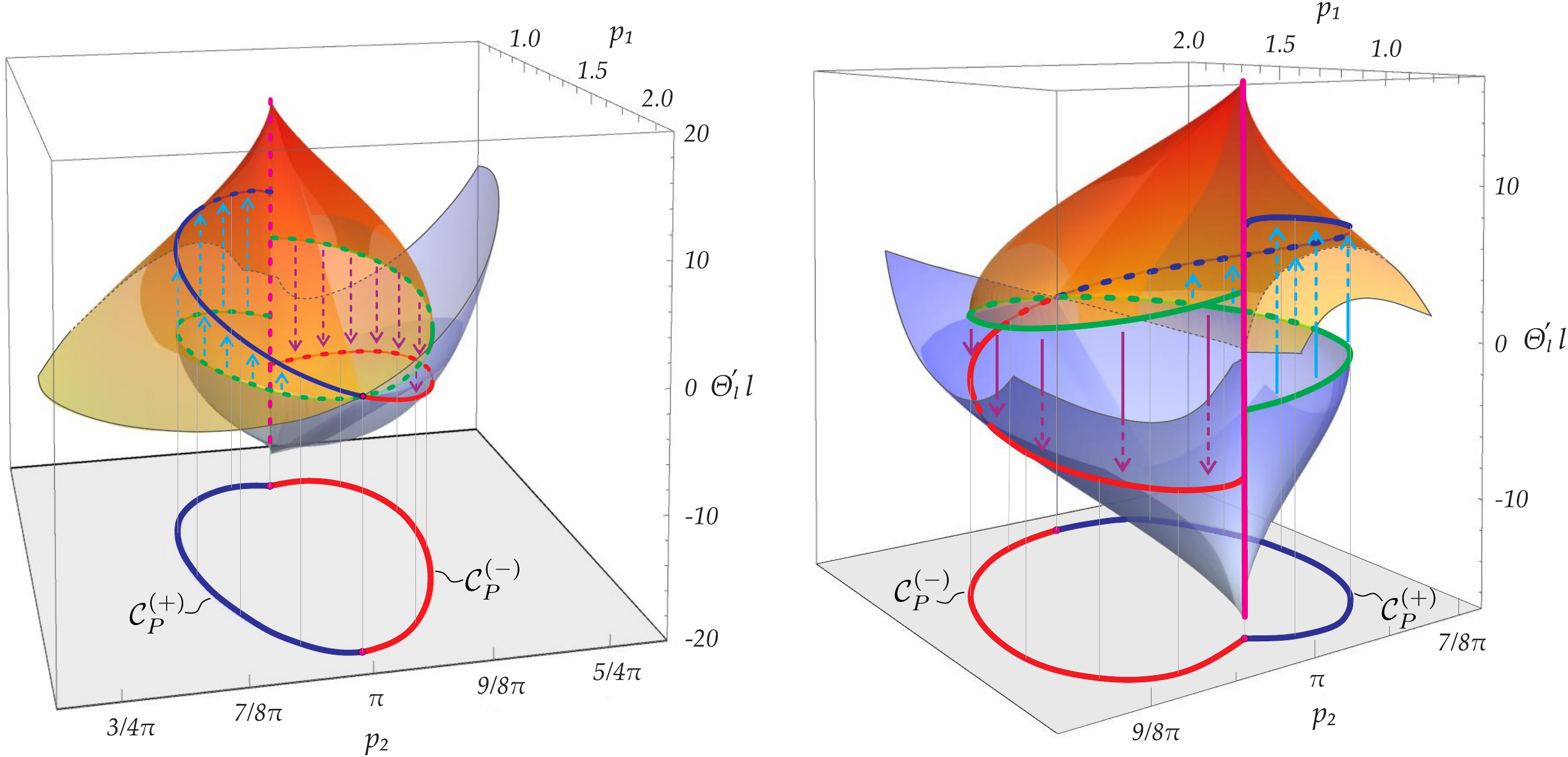}
	\end{center}
	\caption{\footnotesize{
			Curvature at the final curvilinear coordinate, $\Theta'_l=\Theta'(s=l)$, (made dimensionless by multiplication with the length $l$) as a function of the control parameters
			$p_1$ and $p_2$ for ECM-IIb with $\kappa_D=\lambda_D=\alpha=0$, $\upsilon=\pi$, and $\rho=0.8$ (corresponding to the setting considered in the third column, second line, of Fig. \ref{ECMII_alpha0_nupi}). Two opposite views of the three-dimensional plot are reported on the left and the right. Green line highlights critical configurations for which snap occurs toward a configuration laying along the  blue and red curve having the  same control parameters pair but opposite curvature sign.
			The projection of $\mathcal{C}^{(+)}$ (blue line) and $\mathcal{C}^{(-)}$ (red line) on the $p_1-p_2$ plane is also reported as the representation of the catastrophe locus within the control plane. 
	}}
	\label{3dIIb}
\end{figure}

In analogy with the observations for ECM-I machine, the following new features of the catastrophe sets for ECM-II are displayed:
\begin{itemize}\item \textit{Variable number of bifurcation points.}
The catastrophe sets in Figs. \ref{ECMII_alpha0_nu0}, \ref{ECMII_alpha0_nupi}, \ref{ECMII_R05_alpha} and \ref{ECMII_LDnu1} exhibit a number of bifurcation points ranging from two to five. For instance, the ECM-IIa machines  in Fig.  \ref{ECMII_alpha0_nu0} and the ECM-IIb machines  in Fig. \ref{ECMII_alpha0_nupi} have a catastrophe set with two bifurcation points on the symmetry axis. Moreover, In Fig. \ref{ECMII_LDnu1}, the catastrophe set for the ECM-IIa machine for $\upsilon=\pi/4$ and $\lambda_D=0.2$ (second row and second column) has five bifurcation points. 
\item \textit{Convex measure of the catastrophe locus  $\mathcal{C}_P$}. 
The catastrophe sets reported in the first row of Fig. \ref{ECMII_alpha0_nupi} have $\mathsf{C}\simeq1$ for $\rho=\{0.1,\,0.2,\,0.3\}$ (first three columns) and $\mathsf{C}= 0.9998$ for $\rho=0.5$ (fourth column).
\end{itemize}
The three-dimensional representation in Fig. \ref{3dIIb} shows the curvature at the final curvilinear coordinate as a function of the two control parameters for ECM-IIb with $\kappa_D=\lambda_D=\alpha=0$, $\upsilon=\pi$, and $\rho=0.8$ (corresponding to the setting considered in the third column, second line, of Fig. \ref{ECMII_alpha0_nupi}). The same three-dimensional plot is reported on the left and the right under two opposite perspectives. Multiplicity and uniqueness of equilibrium configuration are highlighted for control parameters pairs respectively inside and outside the closed curve defining the catastrophe locus (projection on the $p_1-p_2$ plane). The jump in the equilibrium configuration, displayed at the catastrophe locus with  arrows, implies a change in the curvature sign (for the same value of control parameters).

Finally, the analysis of ECM-II is complemented by a discussion about the special case of the rigid bar with infinitely large length  reported in Sect. \ref{appB_21} of the Supplementary Material and the suggestion for the initial values of the control parameters $\bp(\tau_0)$ in Sect. \ref{appECM-II} of the Supplementary Material.

\section{The physical realization of the \textit{elastica catastrophe machine}}\label{expsect}

\begin{figure}[!htb]
\begin{center}
\includegraphics[width=0.9\textwidth]{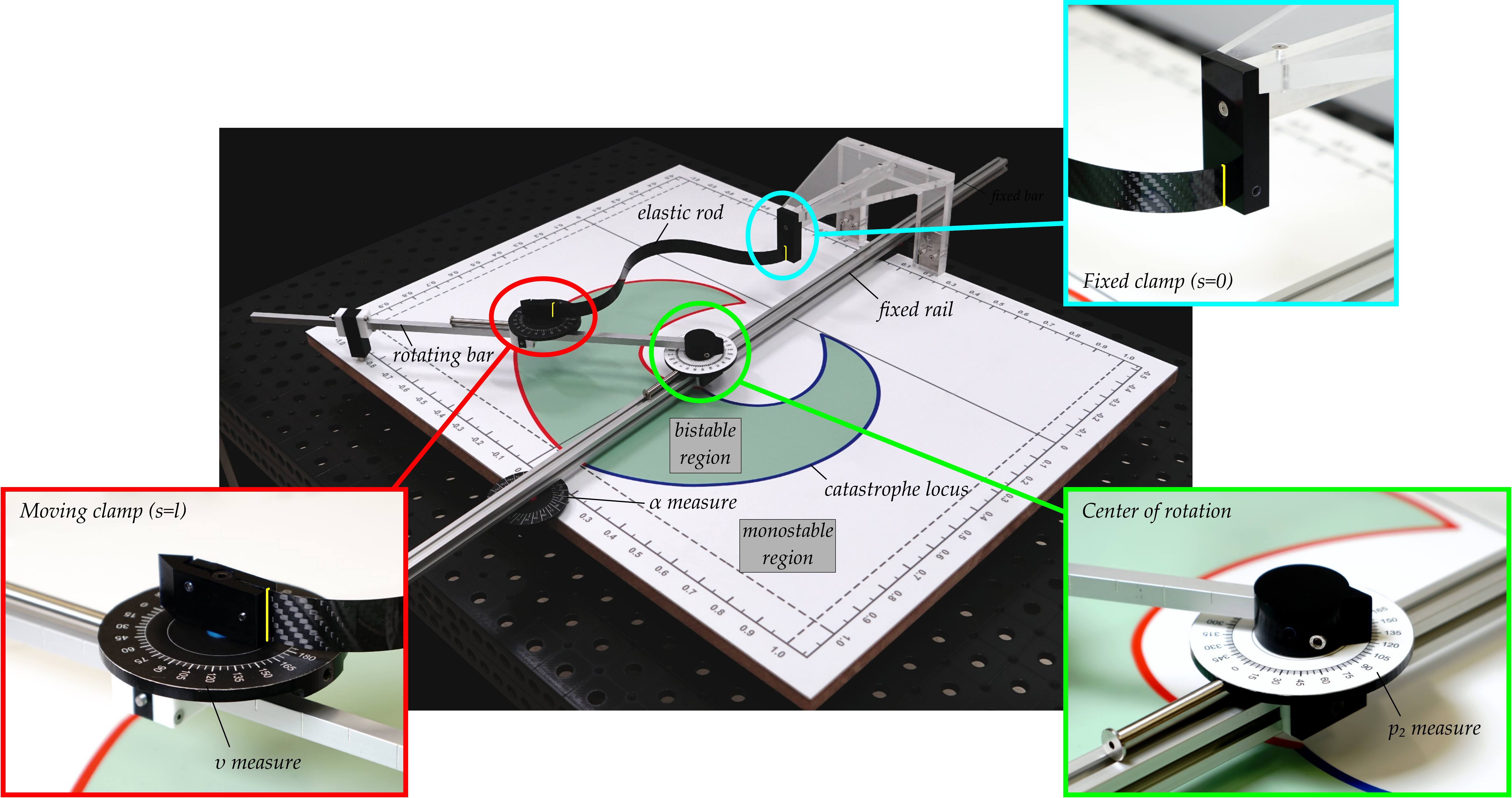}
\end{center}
\caption{\footnotesize{Prototype of the \emph{elastica catastrophe machine}. A (carbon fiber) rod  constrained at its two ends by a fixed clamp (inset upper right, cyan) and another clamp (inset lower left, red) which may move along a stiff bar, which, in turn, may rotate about its end pinned and possibly moving along a fixed rail (inset lower right, green). The  rail is fixed to the plastic desk through two clips. The angles $\upsilon$ and $\alpha$ can be respectively measured from the black goniometers fixed on the movable clamp and the plane, while the angle $p_2$ can be measured from the white goniometer fixed at the rotation centre of the stiff bar.}}
\label{fig_setup}
\end{figure}

A prototype of the \emph{elastica catastrophe machine} was designed and realized at the Instabilities Lab of the University of Trento, Fig. \ref{fig_setup}. This setup, thought to be as versatile as possible, allowed the experimental investigation of both the two proposed families, ECM-I and  ECM-II (a and b). The  two clamps, one fixed and the other moving, constraining the ends of the elastic rod are assembled  on an HDF (High-density fibreboard) desk. This panel acts as a support for mounting the screen printing of the catastrophe locus, which changes by varying the selected design parameters.  More specifically, the  clamp  constraint at the rod coordinate $s=0$ is fixed and mounted on a PMMA structure. The constraint at the other end of the rod ($s=l$) is provided by a clamp which may slide along a rotating aluminium  hollow bar ($10\times 10$ mm  cross-section) with its end pinned to and possibly sliding along an aluminium rail (aluminium extrusions bar, $20\times 20$ mm cross-section), fixed on the desk through two clips. Three goniometers are mounted to measure during the experiments (i.) the angle between the desk and the rail (design angle $\alpha$ in ECM-II), (ii.) the angle between the rail and the rotating bar (to be used for imposing the control angle $p_2$), and (iii.) the angle  between the rotating bar and the moving clamp inclination (design  angle $\upsilon$). 

Each one of the two proposed families can be tested by properly constraining one of the degrees of freedom of the prototype to a fixed value. In particular, ECM-I is attained by fixing the rotating bar end to a specific point $R$ (design parameters $\kappa_R, \lambda_R$) along the fixed rail, while ECM-II by fixing the moving clamp along the rotating bar at a distance $\rho l$ from its end and by defining the inclination $\alpha$ and the passing point $D$ (design parameters $\kappa_D, \lambda_D$) for the fixed rail. 

Rods of (net) length $l=40$ cm with different cross-sections and made up of different materials have been tested. 
In the following, results are shown for two types of rods: a polikristal rod (by Polimark, Young modulus $E=\SI{2750}{MPa}$, density $\rho=\SI{1250}{kg\,m^{-3}}$) with a cross section 20x0.8 mm, and  a carbon fiber rod  (by CreVeR srl, Young modulus $E=\SI{80148}{MPa}$, density $\rho=\SI{1620}{kg\,m^{-3}}$) with a cross section 20x0.45 mm.

\begin{figure}[!htb]
\begin{center}
\includegraphics[width=0.9\textwidth]{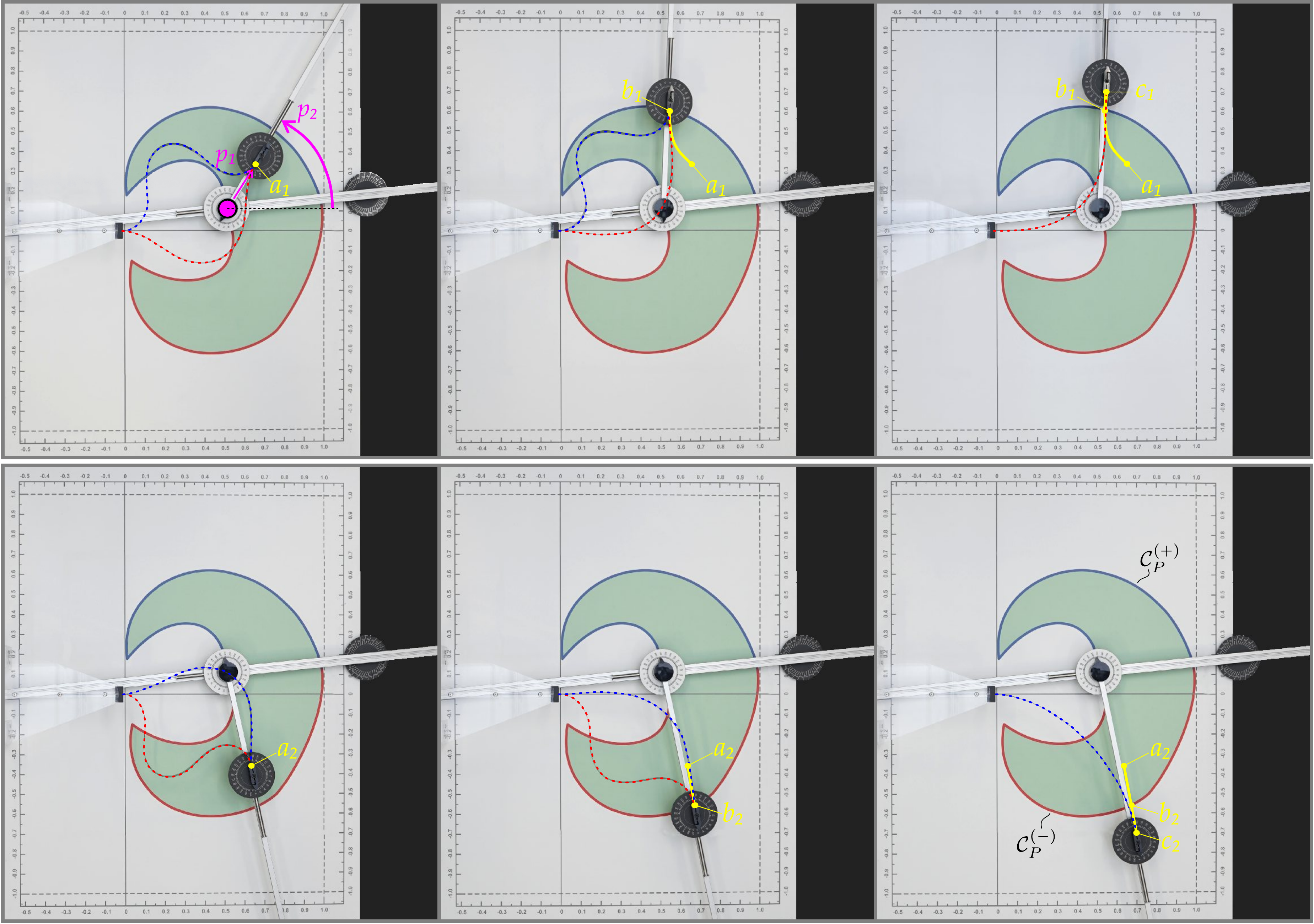}
\end{center}
\caption{\footnotesize{Evolution of the equilibrium configurations for a carbon fiber rod in the ECM-I  (with $\kappa_R=0.5,\,\lambda_R=0.1,\,\upsilon=0$) with varying the position of the moving clamp: within the bistable (green background) region (stage $a$, left column), on the catastrophe locus (stage $b$, central column), and within the monostable region (stage $c$, right column).  The control parameters $\{p_1,\,p_2\}$ are represented in the upper left figure, while deformed configurations with positive/negative curvature at the fixed clamp are highlighted by blue/red dashed curve. At each stage, the clamp position at the previous stage and its path until then is highlighted. While two deformed configurations are possible for stages $a$ and $b$, only one stable configuration exists at stage $c$.}}
\label{fig_frames1}
\end{figure}

The ECM prototype was tested as EMC-I, ECM-IIa, and ECM-IIb for different continuous evolutions of the moving clamp position repetitively crossing  the catastrophe locus and  providing a sequence of snapping mechanisms. Photos taken at specific stages during the experiments are displayed in Fig. \ref{fig_frames1}. In the figure,  all the stable equilibrium configurations are reported at three stages for ECM-I  (with $\kappa_R=0.5,\,\lambda_R=0.1,\,\upsilon=0$). The three stages are related to the position of the moving clamp, located ($a$) within the bistable region, ($b$) on the catastrophe locus, and ($c$) within the monostable region (from left to right in Fig. \ref{fig_frames1}). The two equilibrium configurations related to each of the two first stages (left and central column) differ in the curvature sign at the clamps and are displayed as the superposition of two photos. The only \lq surviving' configuration is displayed at stage $c$ (right column), which can be  reached through a smooth transition from the adjacent deformed configuration or through snapping  from the non-adjacent deformed configuration (characterized by an opposite sign of the curvature at both clamps). The experimental results are reported for a clamp position's evolution ruled by different variations of the control parameters: (i) variation in both the control parameters $p_1$ and $p_2$ (Fig. \ref{fig_frames1}, upper part) and (ii)  variation in the control parameter $p_1$ at fixed value of $p_2$ (Fig. \ref{fig_frames1}, lower part). 

The transition of the deformed configuration during a continuous evolution is highlighted in Figs. \ref{fig_snapshot1} (for ECM-I with $\kappa_R=0.5,\,\lambda_R=0.1,\,\upsilon=0$). 
Four snapshots captured during the fast motion at snapping (taken with a high-speed camera \textit{Sony PXW-FS5}, 120 fps) are superimposed in the second column of the figure (the deformed configurations are highlighted with purple dashed lines), where each two consecutive snapshots are referred to a time interval of approximately 0.15 sec. These sequences display the rod motion towards the change of curvature sign at both clamps. Further experiments performed on different ECM are reported in the section \ref{expsm} of the Supplementary Material.
 
During the experiments, photos were taken with a Sony $\alpha9$ and videos with a high-speed camera (model \textit{Sony PXW-FS5 4K}, 120 fps). A couple of videos showing example of use of the prototype as ECM-I and ECM-IIb are available as supplementary material.\footnote{Movies of experiments can be found in the additional material available at\\
\url{http://www.ing.unitn.it/~dalcorsf/elastica_catastrophe_machine.html}
}

\begin{figure}[!htb]
\begin{center}
\includegraphics[width=0.9\textwidth]{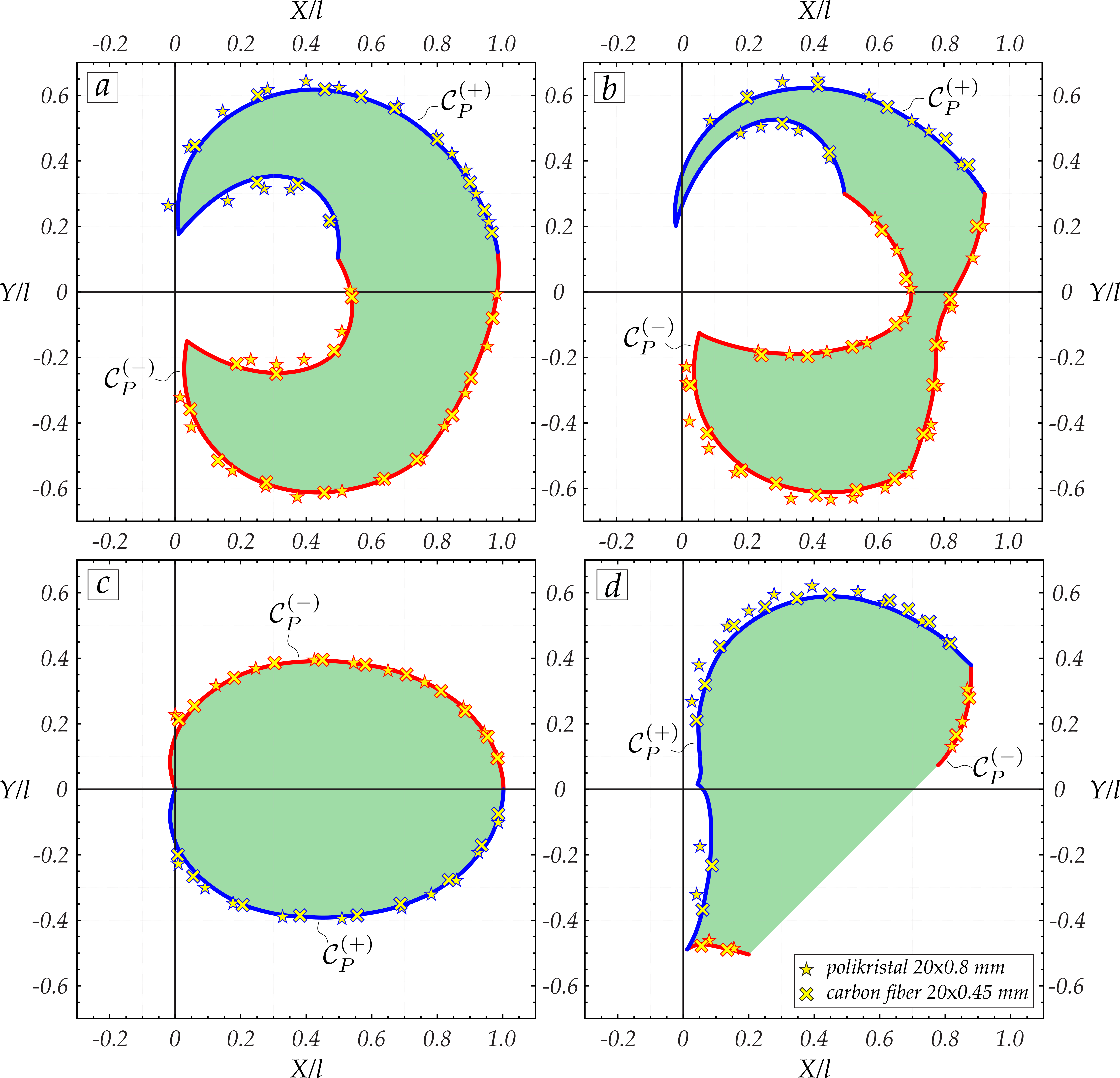}
\end{center}
\caption{\footnotesize{Experimental validation of the theoretically predicted catastrophe locus for ECMs-I ($a$: $\{\kappa_R=0.5,\,\lambda_R=0.1,\,\upsilon=0\}$; $b$: $\{\kappa_R=0.5,\,\lambda_R=0.3,\,\upsilon=0\}$), for ECM-IIb ($c$: $\{\kappa_D=\lambda_D=\alpha=0,\,\rho=1,\,\upsilon=\pi\}$, for ECM-IIa ($d$: $\{\kappa_D=\lambda_D=\upsilon=0,\,\alpha=\pi/4,\,\rho=0.5\}$). Measures from testing polikristal rod of thickness $\SI{0.8}{mm}$ (stars) and with a carbon fiber rod of thickness $\SI{0.45}{mm}$ (crosses) show an excellent agreement with the theoretical predictions.}}
\label{fig_exppoints}
\end{figure}

Finally, the quantitative assessment of the theoretically predicted catastrophe loci is reported in Fig. \ref{fig_exppoints} superimposing the experimental critical points for polikristal (star markers) and carbon fiber (crosses markers) rods. The following settings are shown: (i) ECMs-I (with $\kappa_R=0.5,\,\lambda_R=0.1,\,\upsilon=0$, Fig. \ref{fig_exppoints}a, and with $\kappa_R=0.5,\,\lambda_R=0.3,\,\upsilon=0$, Fig. \ref{fig_exppoints}b), (ii) ECM-IIb (case $\kappa_D=\lambda_D=\alpha=0,\,\rho=1,\,\upsilon=\pi$, Fig. \ref{fig_exppoints}c) and (iii) ECM-IIa (case $\kappa_D=\lambda_D=\upsilon=0,\,\rho=0.5,\,\alpha=\pi/4$, Fig. \ref{fig_exppoints}d). 
The comparisons reported in the figure fully display the experimental validation of the theoretical catastrophe loci. During the tests, very small portions of the catastrophe locus were not investigated because of some unavoidable physical limitations (for instance rod's self-intersection). The  accuracy in the experimental measure of the critical conditions is observed to be higher when using carbon fiber rods. The inferior accuracy  in testing with polikristal rods is expected to be related to the intrinsic viscosity and weight-stiffness ratio of the material.

\section{Conclusions}

For the first time, the design  and the experimental validation of a catastrophe machine has been addressed for a system made up of a continuous and elastic flexible element, extending the classical formulation for discrete systems.
A theoretical framework referring to  primary kinematical quantities and exploiting the concept of the universal snap surface has been introduced. Among the infinite set of  \emph{elastica catastrophe machines}, two families have been proposed and the related catastrophe locus investigated to explicitly show the features of the present model. A parametric analysis has disclosed substantial differences in the shape of the catastrophe locus in comparison with those deriving from classical catastrophe machines. 
In particular, the proposed machines can fulfil peculiar geometrical properties as convexity and a variable number of bifurcation points for the catastrophe loci.
These meaningful characteristics may enhance  the efficiency of snapping devices exploting high-energy release points otherwise unreachable. 
The research has been completed by the  validation of the theoretical results through the physical realization of a prototype enabling experiments with each of the two presented families of \emph{elastica catastrophe  machines}.

\vspace*{5mm} \noindent {\sl Acknowledgments} AC and DM  gratefully
acknowledge financial support from the grant ERC advanced grant \lq
Instabilities and nonlocal multiscale modelling of materials'
ERC-2013-AdG-340561-INSTABILITIES (2014-2019). FDC  gratefully acknowledges financial support from the European Union’s Horizon 2020 research
and innovation programme under the Marie Sklodowska-Curie grant agreement ‘INSPIRE - Innovative
ground interface concepts for structure protection’ PITN-GA-2019-813424-INSPIRE. The authors are grateful to Mr. Lorenzo P. Franchini for the assistance during the experiments and to Prof. Claudio Fontanari (University of Trento) for a stimulating discussion.

\end{document}